\documentclass[11pt]{article}
\usepackage{jheppub}

\usepackage{epsfig}
\usepackage{amssymb}
\usepackage{amsmath}

\usepackage{graphicx, subfigure, array, placeins, float}

\DeclareMathOperator{\Tr}{Tr}
\DeclareMathOperator{\pslash}{\displaystyle{\not}}

\title{Two-Loop QCD Corrections to the Higgs plus three-parton amplitudes with Top Mass Correction}

\author[a,b]{Qingjun Jin}
\emailAdd{qjin@gscaep.ac.cn}
\author[a,c]{and Gang Yang}
\emailAdd{yangg@itp.ac.cn}
\affiliation[a]{CAS Key Laboratory of Theoretical Physics, Institute of Theoretical Physics, \\Chinese Academy of Sciences, Beijing 100190, China}
 \affiliation[b]{Graduate School of China Academy of Engineering Physics, No. 10 Xibeiwang East Road, Haidian District, Beijing, 100193, China}
\affiliation[c]{School of Physical Sciences, University of Chinese Academy of Sciences,  \\No. 19A Yuquan Road, Beijing 100049, China}

\abstract{
We obtain the two-loop QCD corrections to the Higgs plus three-parton amplitudes with dimension-seven operators in Higgs effective field theory. This provides the two-loop S-matrix elements for Higgs plus one-jet production at the LHC with top-mass correction. We apply efficient unitarity plus IBP methods which are described in detail. We also study the color decomposition of the fermion cuts and find a connection between fundamental and adjoint representations which can be used to reduce non-planar to planar unitarity cuts in the Higgs to three-gluon amplitudes. We obtain final results in simple analytic form which exhibits intriguing hidden structures. The principle of maximal transcendentality is found to be satisfied for all results. The lower transcendentality parts also contain universal building blocks and can be written in compact analytic form, suggesting further hidden structures.
}

\begin{document}

\maketitle

\setcounter{footnote}{0}

\section{Introduction}

Scattering amplitudes play an indispensable role in particle physics. They act as a bridge between theory and experiment. The Large Hadron Collider (LHC) verified the correctness of the standard model of particle physics with the discovery of its last missing particle, the Higgs boson \cite{Aad:2012tfa, Chatrchyan:2012xdj}. 
One main object of the present and future collider experiments is to understand more precisely the Higgs properties and the mechanism of electroweak symmetry breaking.
The proposed future colliders, such as the circular electron-positron collider (CEPC) in China \cite{CEPCStudyGroup:2018rmc, CEPCStudyGroup:2018ghi} and the future circular collider (FCC) at CERN \cite{Abada:2019lih, Abada:2019zxq, Benedikt:2018csr}, are expected to provide experimental data with unprecedented precision. 
In order to compare with the experiment, one needs to compute scattering amplitudes to the next-to-next-to leading order (NNLO) or even higher orders. 
This is usually beyond the capability of traditional Feynman diagram methods.
Fortunately, during last thirty years, in the field of amplitudes calculation, many new methods and tools have been developed, including the spinor helicity formalism \cite{Xu:1986xb, DeCausmaecker:1981jtq, Berends:1981rb, Kleiss:1985yh},  the unitarity cut method \cite{Bern:1994zx, Bern:1994cg, Britto:2004nc}, and recursion relations \cite{Berends:1987me, Britto:2004ap, Britto:2005fq}. These methods have achieved great success in the computation of scattering amplitudes, not only in supersymmetric field theories, but also in realistic QCD.

In this paper, we study the Higgs plus three-parton amplitudes in standard model. Our motivation is twofold. 
First, as mentioned above, the precise theoretical prediction of Higgs scattering process is highly demanded to match the improving precision of experiments. 
At the LHC, the dominant Higgs production channel is the gluon fusion through a top quark loop \cite{Ellis:1975ap, Georgi:1977gs}.
The computation of this process can be simplified using an effective field theory (EFT) in which the top quark is integrated out \cite{Wilczek:1977zn, Shifman:1979eb, Dawson:1990zj, Djouadi:1991tka, Kniehl:1995tn, Chetyrkin:1997sg, Chetyrkin:1997un}. This EFT is a good approximation when the top mass $m_t$ is much larger than Higgs mass $m_H$. 
The leading term in the effective Lagrangian is a unique dimension-5 operator, $H{\rm tr}(F_{\mu\nu}F^{\mu\nu})$, where $H$ is the Higgs field and $F_{\mu\nu}$ is the gauge field strength.
The two-loop QCD corrections to Higgs plus three-parton amplitudes with the leading dimension-5 operator were computed in \cite{Gehrmann:2011aa},
which have been used to obtain the cross sections of Higgs plus a jet production at N$^2$LO \cite{Boughezal:2013uia,Chen:2014gva,Boughezal:2015aha,Boughezal:2015dra, Harlander:2016hcx, Anastasiou:2016hlm, Chen:2016zka} in the infinite top mass limit.
When the Higgs transverse momentum is comparable to the top mass, the contribution of higher dimension operators in the Higgs EFT will be important. 
This has been taken into account so far only at NLO QCD accuracy, including the finite top mass effect \cite{Lindert:2018iug,Jones:2018hbb,Neumann:2018bsx}.
A concrete goal of this paper is to compute the two-loop QCD corrections for Higgs plus 3-parton amplitudes with dimension-7 operators in the Higgs EFT. This provides, for the first time at N$^2$LO QCD accuracy, the S-matrix elements of the top mass correction for Higgs plus a jet production.

Another motivation is to study the hidden analytic structures of amplitudes.
One particular focus of this paper is related to the so-called maximal transcendentality principle (MTP). Transcendentality is a mathematical quantity used to characterize the algebraic complexity of functions or numbers. 
The principle of maximal transcendentality conjectures that the algebraically most complicated part of certain physical observables in QCD and $\mathcal{N}=4$ SYM are equal.
It was first proposed in \cite{Kotikov:2002ab, Kotikov:2004er} that, the anomalous dimensions of twist-two operators in ${\cal N}=4$ SYM can be obtained from the maximally transcendental part of the QCD results  \cite{Moch:2004pa}. 
Intriguingly, the principle can be extended to the Higgs plus three-parton amplitudes or form factors, which involes complicated two dimensional Harmonic Polylogarithms \cite{Gehrmann:2000zt, Gehrmann:2001jv}.
This was first observed for the $\Tr(F^2)\rightarrow 3g$ form factors \cite{Brandhuber:2012vm}, which on one side correspond to the QCD corrections to the Higgs to 3-parton amplitudes in the infinite top quark mass limit \cite{Gehrmann:2011aa}, and on the other side is equivalent to the form factors of stress-tensor multiplet in ${\cal N}=4$ SYM.
(The universal property of the maximally transcendental parts were also found in form factors of more general operators in ${\cal N}=4$ SYM \cite{Brandhuber:2014ica, Loebbert:2015ova, Brandhuber:2016fni, Loebbert:2016xkw, Banerjee:2016kri}.) Besides, the MTP was verified for other quantities like Wilson lines \cite{Li:2014afw, Li:2016ctv}, and has also been applied to compute collinear anomalous dimensions \cite{Dixon:2017nat}. 
Recently, the MTP was found to be true for three-gluon form factors of the the dimension-6 operators in the pure gluon sector \cite{Brandhuber:2017bkg, Jin:2018fak, Brandhuber:2018xzk, Brandhuber:2018kqb}. 
In this paper, as also reported in \cite{Jin:2019ile}, we show that the MTP can be extended to the form factors with external fundamental quarks: with a simple replacement of the quadratic Casimirs $C_F\rightarrow C_A$, the maximally transcendental (MT) part of $H\rightarrow q\bar{q}g$ form factors were found to reduce to the MT part of $H\rightarrow 3g$ form factors. 

Furthermore, we find that  the lower transcendentality parts also exhibit certain universality.
For example, the transcendentality degree-3 parts can be constructed by a building block $T_3$, plus simple log functions and constants. The degree-2 parts also contain building blocks $T_2$ and $T'_2$. With these building blocks, the amplitudes can be written in compact forms,
which suggest that hidden structures also exist in the lower transcendental parts. Exploring these structures further will be important for the computation of full QCD amplitudes.

Our computations employ a new strategy of combining unitarity cut \cite{Bern:1994zx, Bern:1994cg, Britto:2004nc} and integration by parts (IBP) methods \cite{Chetyrkin:1981qh, Tkachov:1981wb}. 
We apply IBP directly to the cut integrands, which are constructed using unitarity cuts. This avoids the non-trivial reconstruction of the full integrand.  
The strategy also increases the efficiency of IBP reduction significantly, with the cut constraints imposed.  
Similar strategy has also been used in \cite{Boels:2018nrr}. Ideas of  applying cuts to simplify IBP reductions were also studied in \cite{Kosower:2011ty, Larsen:2015ped, Ita:2015tya, Georgoudis:2016wff, Abreu:2017xsl, Abreu:2017hqn}.
The pure gluon sector of two-loop $H\rightarrow 3g$ amplitudes contains only the leading color contribution, in which the loop integrands can be conveniently obtained using the planar unitarity method. In the presence of internal quarks, more complicated color structures appears. We will show that by making connection between fermions in fundamental and adjoint representations, a color decomposition is possible such that the full two-loop $H\rightarrow 3g$ integrand can be constructed using planar cuts.

This paper is organized as follows. In Section \ref{sec:setup}, we review the Higgs EFT and describe the divergence structures. In Section \ref{sec:computation}, we describe the details of the computation using unitarity-IBP strategy. In Section \ref{sec:color-decomp}, we discuss the color decomposition of amplitudes that involve internal quarks. In Section \ref{sec:results}, we present the analytic results of form factors. We conclude and discuss the transcendentality properties in Section \ref{sec:discussion}. Appendices \ref{app:oneloop}--\ref{app:remainderO4} provide the expressions of one-loop and two-loop results.

\section{Preparations}
\label{sec:setup}

In this section, we first describe the Higgs effective action and the dimension seven operators, then we review the subtraction of divergences and provide explicit formulae. 

\subsection{Operator basis}
\label{sec:operators}

The Higgs boson can be produced from the gluon fusion through a heavy quark loop at the LHC. The Yukawa couplings between Higgs and quarks are proportional to the mass of quarks, so the  diagrams with a top quark loop dominate. Integrating out the top quark renders the Higgs effective field theory (HEFT) \cite{Wilczek:1977zn, Shifman:1979eb, Dawson:1990zj, Djouadi:1991tka, Kniehl:1995tn, Chetyrkin:1997sg, Chetyrkin:1997un}:
\begin{equation}
{\cal L}_{\rm eff} = \hat{C}_0H \mathcal{O}_0 + {1\over m_{\rm t}^2} \sum_{i=1}^4 \hat{C}_i H \mathcal{O}_i + {\cal O}\left( {1\over m_{\rm t}^4} \right) \,,
\label{eq:HiggsEFT}
\end{equation}
where $\hat{C}_i$ are Wilson coefficients, $H$ is the Higgs field, ${\cal O}_0 = {\rm tr}(F^2)$ is the leading term, and the subleading terms contain dimension-6 operators \cite{Buchmuller:1985jz, Gracey:2002he, Neill:2009tn, Harlander:2013oja, Dawson:2014ora}:
\begin{align}
\mathcal{O}_{1} & = {\rm tr}(F_\mu^{~\nu} F_\nu^{~\rho} F_\rho^{~\mu}) \,, \\ 
\mathcal{O}_{2} & = {\rm tr}(D_\rho F_{\mu\nu} D^\rho F^{\mu\nu} ) \,,\\
\mathcal{O}_{3} & = {\rm tr}(D^\rho F_{\rho\mu} D_\sigma F^{\sigma\mu}) \,,\\
\mathcal{O}_{4} & = {\rm tr}(F_{\mu\rho} D^\rho D_\sigma F^{\sigma\mu}) \,.
\end{align}
The last two operators have zero contribution in the pure gluon sector and only contribute when there are internal quark lines. In this paper we will consider the full QCD corrections including massless quarks contributions.

An amplitude with a Higgs boson and $n$ gluons is equivalent to the form factor with an operator ${\cal O}_i$ in the EFT \eqref{eq:HiggsEFT}:
\begin{equation}
{\cal F}_{{\cal O}_i,n} = \int d^4 x \, e^{-i q\cdot x} \langle p_1, \ldots, p_n | {\cal O}_i(x) |0 \rangle \,,
\end{equation}
where $q^2 = m_H^2$. 
In the following, we will often refer Higgs amplitudes as form factors.

Using Bianchi identity one can decompose the operator ${\cal O}_2$ as (see e.g. \cite{Gracey:2002he})
\begin{align}
{\cal O}_{2} = {1\over2}\, \partial^2{\cal O}_{0} -4\, g_{\textrm{\tiny YM}} \, {\cal O}_{1} +2\, {\cal O}_{4} \,.
\label{eq:ope-linear-relation}
\end{align}
The operator relation can be transformed into a relation for the form factors:
\begin{align}
\label{eq:O1-linear-relation}
{\cal F}_{{\cal O}_{2}} = {1\over2}\, q^2 \,{\cal F}_{{\cal O}_{0}} -4\, g_{\textrm{\tiny YM}} \, {\cal F}_{{\cal O}_{1}} +2\, {\cal F}_{{\cal O}_{4}} \,,
\end{align}
where the partial derivatives reduce to square of $q$ which is the total momentum flowing through the ${{\cal O}_0}$ operator. 
This relation can serve as a self-consistency check for our computations.

One can classify the operators according to their {\it length}. Naturally, the length of an operator $\mathcal{O}$ is the number of elementary fields ($A$, $\bar{\psi}$ and $\psi$) in its lowest expansion (i.e.~with minimal number of elementary fields). For example ${\rm tr}(F^2)\sim {\rm tr}(\partial^2A^2)$ has length 2, and ${\rm tr}(F^3)\sim {\rm tr}(\partial^3A^3)$ has length 3. A form factor of an operator is called ``minimal" if the form factor contains exactly the same number of on shell particles as that of the lowest expansion of the operator.  For example ${\rm tr}(F^3)\rightarrow ggg$ and $\epsilon_{ijk}\psi^i\psi^j\psi^k \rightarrow qqq$ are minimal form factors, but ${\rm tr}(F^2)\rightarrow ggg$ is a non-minimal form factor.

Sometimes this ``naive" definition results in a zero minimal tree form factor.  As an example, for $\mathcal{O}_4$ the tree form factor with two external gluons vanishes, and the simplest non-zero tree form factor is  $\mathcal{O}_4\rightarrow q\bar{q}g$. The reason is that, using the equation of motion $D_\sigma F^{\sigma\mu}\sim g \sum_{i}(\bar\psi_i\gamma^\nu T^a \psi_i)$, $\mathcal{O}_4$ is equivalent to  ${\cal O}_4' =F_{\mu\nu}^a D^\mu \sum_{i}(\bar\psi_i\gamma^\nu T^a \psi_i)$, which is a length-3 operator. 
The more proper definition is that, the minimal form factor for a given operator is the simplest form factor which is non-zero at tree level, and the length of the operator is the number of external on-shell states in the minimal form factor. Using this definition, ${\cal O}_4$ has length 3,  and its minimal form factor is $\mathcal{O}_4\rightarrow q\bar{q}g$. Similarly, $\mathcal{O}_3$ has length 4, and its minimal form factor is $\mathcal{O}_3\rightarrow qq\bar{q}\bar{q}$.

\subsection{Divergence structures}
\label{sec:divergence-structure}

Form factors contain both UV and IR divergences. We apply dimensional regularization ($D=4-2\epsilon$) in the conventional dimension regularization (CDR) scheme, and for the renormalization, we use the modified minimal subtraction renormalization ($\overline{\rm MS}$) scheme \cite{Bardeen:1978yd}. To subtract the IR divergences, we apply the Catani subtraction formula  \cite{Catani:1998bh}. Below we describe these in detail.

To begin with, the bare form factor can be expanded as
\begin{align}\label{fbare}
{\cal F}_{\rm b} = g_0^{\delta_n} \left[ {\cal F}_{\rm b}^{(0)} + {\alpha_0 \over 4\pi} {\cal F}_{\rm b}^{(1)} + \Big( {\alpha_0 \over 4\pi} \Big)^2 {\cal F}_{\rm b}^{(2)} + {\cal O}(\alpha_0^3) \right] \,,
\end{align}
where $g_0 = g_{\textrm{\tiny YM}}$ is the bare gauge coupling and $\alpha_0 = \frac{g_0^2}{4\pi}$. We pull out the coupling $g_0^{\delta_n}=g_0^{n-\mathbb{L}}$ in the tree form factor, which depends on the number of external legs $n$ and the length of the operator $\mathbb{L}$. 

The renormalization of the UV divergences can be implemented in two steps, one for the coupling constant and one for the local operator. 

First, we express the bare coupling $\alpha_0$ in terms of the renormalized coupling $\alpha_s =\alpha_s(\mu^2) = \frac{g_s(\mu^2)^2}{4\pi}$, evaluated at the renormalization scale $\mu^2$, as
\begin{align}
\alpha_0 & = \alpha_s  S_\epsilon^{-1} {\mu^{2\epsilon} \over \mu_0^{2\epsilon}}  \Big[ 1 - {\beta_0 \over \epsilon} {\alpha_s \over 4\pi} + \Big( {\beta_0^2 \over \epsilon^2} - {\beta_1 \over 2 \epsilon} \Big) \Big({\alpha_s \over 4\pi}\Big)^2 + {\cal O}(\alpha_s^3) \Big] \,, 
\end{align}
where the factor $S_\epsilon = (4\pi e^{- \gamma_{\text{E}}})^\epsilon$ is due to the use of ${\overline {\rm MS}}$ scheme, and $\mu_0^2$ is the scale introduced to keep gauge coupling dimensionless in the bare Lagrangian. 
The first two coefficients of the $\beta$ function are\footnote{Since in our result $n_f$ is always associated with a factor $t_F$ in $\Tr(T^aT^b)=t_F\delta^{ab}$, we simply set $t_F=1/2$.}
\begin{align}
\beta_0 = {11 C_A \over 3} - {2 n_f \over 3} \,, \qquad \beta_1 = {34 C_A^2 \over 3} - {10 C_A n_f \over 3} - 2 C_F n_f \,,
\end{align}
where $n_f$ is the flavor number of fermions and $C_A$ and $C_F$ are the quadratic Casimirs in the adjoint and fundamental representations:
\begin{align}
C_A = N_c \,, \qquad C_F = \frac{N_c^2 -1}{2N_c} \,.
\end{align}

Second, we renormalize the operator by introducing the renormalization constant  $Z$ for the operator
\begin{equation}
Z = 1 + \sum_{l=1}^\infty \Big({\alpha_s \over 4\pi}\Big)^l Z^{(l)}  \,.
\label{eq:Z_def_expand}
\end{equation}
The anomalous dimension can be computed from the renormalization constant as 
\begin{align}
\gamma &= \mu \frac{\partial}{\partial \mu} \log Z  = \sum_{l=1}^\infty \left(\frac{\alpha_s}{4\pi}\right)^l \gamma^{(l)} \,.
\end{align}
Using \eqref{eq:Z_def_expand} and note that $\mu \frac{\partial}{\partial \mu} \alpha_s(\mu) = - 2 \epsilon \alpha_s - {\beta_0 \over 2\pi} \alpha_s^2 + {\cal O}(\alpha_s^3)$, we have
\begin{align}
\gamma^{(1)} & = {2\epsilon} Z^{(1)} \,, \\
\gamma^{(2)} & = {4\epsilon} Z^{(2)} - {2\epsilon} \big(Z^{(1)} \big)^2 + 2 Z^{(1)} \beta_0 \,.
\label{eq:ADfromZ-2loop}
\end{align}
Since $\gamma$ is finite, it is clear that the  ${1\over\epsilon^2}$ part in $Z^{(2)}$ is fixed by the one-loop results as
\begin{equation}
Z^{(2)} \big|_{{1\over\epsilon^2}\textrm{-part.}} = {1\over2} \big(Z^{(1)} \big)^2 - {1\over2 \epsilon} Z^{(1)} \beta_0 \,.
\label{eq:Z2loop_eps2}
\end{equation}

Expanding the renormalized form factor as 
\begin{align}
{\cal F} \equiv Z \,{\cal F}_{\rm b}= g_s^{\delta_n} \, S_\epsilon^{-\delta_n/2} \sum_{l=0}^\infty \Big( {\alpha_s \over 4\pi} \Big)^l {\cal F}^{(l)}  \,,
\end{align}
we have the relations between the renormalized components ${\cal F}^{(l)}$ and the bare ones ${\cal F}_{\rm b}^{(l)}$ as
\begin{align}
{\cal F}^{(0)} & = {\cal F}_{\rm b}^{(0)} \,, \\
{\cal F}^{(1)} & = S_\epsilon^{-1} {\cal F}_{\rm b}^{(1)} +  \Big( Z^{(1)} - {\delta_n\over2} {\beta_0 \over \epsilon} \Big) {\cal F}_{\rm b}^{(0)}  \,, 
\label{eq:F1loopIR} \\
{\cal F}^{(2)} & = S_\epsilon^{-2} {\cal F}_{\rm b}^{(2)} + {S_\epsilon^{-1}} \Big[ Z^{(1)} - \Big(1+{\delta_n\over2} \Big) {\beta_0 \over \epsilon} \Big] {\cal F}_{\rm b}^{(1)} \nonumber\\
& +  \Big[ Z^{(2)} - {\delta_n\over2} {\beta_0 \over \epsilon} Z^{(1)} + {\delta_n^2+2\delta_n\over8} {\beta_0^2 \over \epsilon^2} - {\delta_n\over4} {\beta_1 \over \epsilon} \Big] {\cal F}_{\rm b}^{(0)} \,.
\label{eq:F2loopIR} 
\end{align}

The renormalized form factors contain only IR divergences, which take a universal structure \cite{Catani:1998bh, Sterman:2002qn} (see also \cite{Gehrmann:2011aa}): 
\begin{align}
{\cal F}^{(1)} &= I^{(1)}(\epsilon) {\cal F}^{(0)} + {\cal F}^{(1),{\rm fin}} + {\cal O}(\epsilon) \,,  \\
{\cal F}^{(2)} &= I^{(2)}(\epsilon) {\cal F}^{(0)} +  I^{(1)}(\epsilon) {\cal F}^{(1)} + {\cal F}^{(2),{\rm fin}} + {\cal O}(\epsilon)   \,,
\end{align}
where for the form factor with three external gluons, we have
\begin{align}
I_{3g}^{(1)}(\epsilon) &= - {e^{\gamma_E \epsilon} \over \Gamma(1-\epsilon)} \bigg( \frac{C_A}{\epsilon^2} + \frac{\beta_0}{2 \epsilon} \bigg) \sum_{i=1}^n (-{s_{i,i+1}} )^{-\epsilon} \,, \\
I_{3g}^{(2)}(\epsilon) &= - {1\over2} \big[ I^{(1)}(\epsilon) \big]^2   -  {\beta_0 \over \epsilon} I^{(1)}(\epsilon) 
 + {e^{-\gamma_E \epsilon} \Gamma(1-2\epsilon) \over \Gamma(1-\epsilon)} \left[ \frac{\beta_0}{\epsilon} + {\cal K}\right] I^{(1)}(2\epsilon)  \nonumber
+ n {e^{\gamma_E \epsilon} \over \epsilon \Gamma(1-\epsilon)}{\cal H}_{\Omega,g}^{(2)}  \,. \nonumber
\end{align}
For the case with external quarks, we have
\begin{align}
I_{q\bar{q}g}^{(1)}(\epsilon) &= - {e^{\gamma_E \epsilon} \over \Gamma(1-\epsilon)} \bigg[ \bigg( \frac{C_A}{\epsilon^2} + \frac{3C_A}{4\epsilon} + \frac{\beta_0}{4 \epsilon} \bigg) \big( (-{s_{13}} )^{-\epsilon} + (-{s_{23}} )^{-\epsilon} \Big) - {1\over C_A} \bigg( \frac{1}{\epsilon^2} + \frac{3}{2\epsilon} \bigg) (-{s_{12}} )^{-\epsilon}  \bigg] \,, \\
I_{q\bar{q}g}^{(2)}(\epsilon) &= - {1\over2} \big[ I^{(1)}(\epsilon) \big]^2   -  {\beta_0 \over \epsilon} I^{(1)}(\epsilon)  + {e^{-\gamma_E \epsilon} \Gamma(1-2\epsilon) \over \Gamma(1-\epsilon)} \left[ \frac{\beta_0}{\epsilon} + {\cal K} \right] I^{(1)}(2\epsilon) + {e^{\gamma_E \epsilon} \over \epsilon \Gamma(1-\epsilon)}{\cal H}_\Omega^{(2)} \,, \nonumber
\end{align}
where
\begin{equation}
{\cal K} = \left({67\over9} - {\pi^2\over3}\right) C_A - {10\over9}n_f \,,
\end{equation}
and
\begin{align}
{\cal H}_\Omega^{(2)} & = 2 {\cal H}_{\Omega,q}^{(2)} + {\cal H}_{\Omega,g}^{(2)} \,, \\
{\cal H}_{\Omega,g}^{(2)} & =  \left( \frac{\zeta_3}{2} + {5\over12} + {11\pi^2 \over 144} \right)C_A^2 + {5 n_f^2 \over 27} - \left( {\pi^2 \over 72} + {89 \over 108} \right) C_A n_f - {n_f\over 4 C_A}  \,, \\
{\cal H}_{\Omega,q}^{(2)} & =  \left( \frac{7\zeta_3}{4} + {409\over864} - {11\pi^2 \over 96} \right)C_A^2 - \left( \frac{\zeta_3}{4} + {41\over108} + {\pi^2 \over 96} \right) - \left( \frac{3\zeta_3}{2} + {3\over 32} - {\pi^2 \over 8} \right) {1\over C_A^2} \nonumber\\
& + \left( {\pi^2 \over 48} - {25 \over 216} \right) { 2 C_F n_f}  \,.
\end{align}

\section{Computation with unitarity-IBP}
\label{sec:computation}

The traditional method of computing scattering amplitudes is based on Feynman diagrams. In multiloop calculations, this traditional method is not very efficient. This is mainly because the gauge symmetry, unitarity and other properties of the scattering amplitude are destroyed, when the amplitude is split into Feynman diagrams.
The modern unitarity method uses tree amplitudes as building blocks to construct the integrand of loop amplitudes \cite{Bern:1994zx, Bern:1994cg, Britto:2004nc}. In this construction, the original properties and symmetry of the amplitude are mostly preserved, so that the integrand can be calculated much more efficiently. 

The commonly used strategy of unitarity method is to first construct the full integrand using a set of unitarity cuts.
The complete amplitude (before integration) contains a set of loop integrals whose coefficients are rational in the momentum invariants and the spacetime dimension $D$. Each unitarity cut can be used to fix some of the coefficients, and different unitarity cuts will be applied successively until all the coefficients are fixed. 
After the full integrand is obtained via unitarity, the integration by parts (IBP) method can then be used to reduce the amplitude further to a small set of master integrals \cite{Chetyrkin:1981qh, Tkachov:1981wb}.
We illustrate the above procedure as:
\begin{equation}
\mathcal{F}^{(l)}\Bigr|_{\rm cut} \xrightarrow{\ \text{unitarity construction}\ }\mathcal{F}^{(l)}=\sum_a C_aI_a\xrightarrow{\ \text{IBP}\ } \sum_i c_iM_i\ ,
\end{equation}
where $M_i$ are the IBP master integrals.

This strategy has two potential drawbacks. First, rebuilding the complete integrand is not a trivial task.
The labelings of loop momenta in different unitarity cuts are usually different from each other. So the reconstruction of the full integrand involves cumbersome shifting and redefinition of loop momenta, especially when non-planar graphs are involved.\footnote{This problem can be avoided in planar graphs by using zone variables.}  Second, the IBP reduction for the full amplitude can be very slow. IBP  usually takes a long time and consumes a lot of computing resources, and it is sometimes the main bottleneck in the whole calculation. 

We use a new strategy of combining unitarity and IBP which helps to overcome both issues above. The key idea is that instead of applying IBP to the full loop amplitude, we apply IBP directly to each cut integrand.
If a master integral allows a given unitarity cut, this cut will be enough to determine the \emph{final} coefficient of the master integral. A single unitarity cut only fixes the coefficients of a subset of master integrals. One needs to apply different unitarity cuts successively, until all the coefficients are fixed.
In this way, there is no need to construct the full integrand, and one obtains the final coefficients $c_i$ of IBP master integrals directly. 
This strategy can be illustrated as:
\begin{equation}
\mathcal{F}^{(l)}\Bigr|_{\rm cut}=\sum_a C_aI_a|_{\rm cut}\xrightarrow{\ \text{unitarity-IBP}\ } \sum_{\text{cut permitted}} c_iM_i\xrightarrow{\ {\rm collect}\ }\sum_i c_iM_i\ .
\end{equation}
Furthermore, imposing the cut condition drops a lot of integrals and makes a lot of sectors trivial during IBP. 
Our strategy typically increased the efficiency of IBP by an order of magnitude. 
A further important bonus of the unitarity-IBP method is that different cuts can provide internal consistency checks, which are very helpful in complicated cases.

The idea of applying cuts to simplify the IBP reduction has also been used in e.g.~\cite{Kosower:2011ty, Larsen:2015ped, Ita:2015tya, Georgoudis:2016wff}. In those cases, the loop integrand is generally taken as a known input. The strategy used here is different in the sense that its main purpose is to simplify the unitarity construction of amplitudes (from the scratch using tree products), while IBP with cut is only one natural step involved. Similar strategy has also been used in the numerical unitarity approach \cite{Abreu:2017xsl, Abreu:2017hqn}, where unitarity cut and IBP are carried out together with numerical momentum variables to avoid large intermediate expressions. Here our approach is purely analytical and does not involve numerical reconstructions.
We will illustrate our strategy with explicit examples later.

\subsection{$D$-dimensional unitary cut}\label{3.1}

Four-dimensional spinor helicity formalism is very powerful in the computation of supersymmetric gauge theory amplitudes. However, in the computation of non-supersymmetric theory amplitudes, it fails to capture the rational terms (see e.g. \cite{Bern:2007dw, Xiao:2006vr, Ossola:2008xq}). 
We will apply the more general $D$-dimensional unitarity method in the computation of $H\rightarrow 3g$ amplitudes. In this case, it is also enough to consider only the planar cuts. The building blocks are color-stripped tree amplitudes and form factors, which can be computed using planar Feynman diagrams, or recursion relations \cite{Berends:1987me, Britto:2004ap, Britto:2005fq}.
The polarization vectors of cut internal gluons  satisfy the following contraction rule 
\begin{equation}
\varepsilon^{\mu}(p) \circ \varepsilon^{\nu}(p)\equiv \sum_{\rm helicities}\varepsilon^{\mu}(p) \varepsilon^{\nu}(p) =\eta^{\mu\nu}
-\frac{q^{\mu}p^{\nu}+q^{\nu}p^{\mu}}{q\cdot p} \,,
\label{eq:helicity-contraction-rule}
\end{equation}
where $q^\mu$ is an arbitrary reference momenta. The $q$-dependent terms vanish due to gauge invariance, and disappear in the full cut-amplitude.
The quark (or anti-quark) fields also have two external states, denoted by $u_s(p)$ (or $\bar{u}_s(p)$), which are the solutions of the (massless) Dirac equation. 
The contraction rule of internal quark states is
\begin{equation}
u_s(p) \circ \bar{u}_s(p) \equiv \sum_s u_s(p)\bar{u}_s(p)=\pslash{p} \,. 
\label{eq:helicitysum-fermion}
\end{equation}

Comparing with the four-dimensional unitarity cut in spinor helicity formalism, the $D$-dimensional unitarity method usually generates much larger expressions in the intermediate steps. As a compensation, the $D$-dimensional unitarity method not only captures all rational-type terms, but also produces integrals with regular propagators, which is ready for the IBP reduction. In contrast, in the case of four-dimensional unitarity cut, a reconstruction must be performed to convert the spinor-brackets to standard propagators.

In the pure gluon sector the non-planar contribution of $H\rightarrow 3g$ amplitudes vanishes at two loops \cite{Jin:2018fak} and the amplitudes are proportional to the simple color factor $N_c^2$, so the planar unitarity cut gives the full result.  In the presence of internal quark legs, the amplitudes contain $N_c^0$ and $N_c^{-2}$ contributions. However, as will be shown in Section \ref{sec:color-decomp}, we can still use planar cuts,  if we assign proper color factors to different internal-state configurations. So these contributions are not intrinsically non-planar.

The planar unitarity cuts (with color-stripped amplitudes as input) are not suffice in the computation of $H\rightarrow q\bar{q}g$ amplitudes, which contain intrinsic non-planar contributions. 
One may carry out the non-planar unitarity cut, in which the building blocks will be the amplitudes with full color factors.
However, unlike the planar diagrams which can be written in unique forms with the help of zone variables, in the non-planar cut, the same integral may appear in different forms which are related by shifting loop momenta. Extra amount of work needs to be done to bring these different copies of integrals to the canonical forms, and to compare the results of different unitarity cuts. A naive application of this strategy typically makes the non-planar unitarity method less efficient.
Instead, we have computed these non-planar contributions using standard Feynman diagrams (with FeynArts \cite{Hahn:2000kx}) plus IBP reduction methods.
 It would be very desirable to develop an efficient way to apply unitarity cut method in the non-planar sector which we leave for future work.

\subsection{Gauge invariant basis}

The unitarity cut integrand is explicitly gauge invariant, since all its tree building blocks are gauge invariant. This means the cut-integrand vanishes if any $\varepsilon_i\rightarrow p_i$, even before IBP is performed. This explicit gauge invariance serves as a self-consistency check of our cut-integrand.  By contrast, the complete \emph{uncut}-loop integrand is typically not explicitly gauge invariant, setting $\varepsilon_i\rightarrow p_i$ leaves some scaleless integrals which are zero only after integration. 

Since the (cut) amplitude is gauge invariant, we can expand it using a set of gauge invariant basis $B_{\alpha}$ (see e.g. \cite{Gehrmann:2011aa} and also \cite{Boels:2017gyc, Boels:2018nrr} for recent general discussion):\footnote{The expansion in gauge invariant basis applies to both cut and full amplitudes, so we do not distinguish them.}
\begin{equation}
{\cal F}_{n}(\varepsilon_i,p_i,l_a)=\sum_{\alpha} {f}^{\alpha}_{n}(p_i, l_a)B_{\alpha}  \,.
\label{eq:FprojectedinBandf}
\end{equation}
The coefficients ${f}^{\alpha}_{n}(p_i, l_a)$ can be computed as
\begin{equation}
{f}^{\alpha}_{n}(p_i, l_a)=B^{\alpha}\circ{\cal F}_{n}(\varepsilon_i,p_i,l_a) \,,
\end{equation}
where the dual basis $B^{\alpha}$ play as projectors, which satisfies,
\begin{equation}
B^{\alpha}\circ B_{\beta}=\delta^{\alpha}_{\beta}\,, \qquad
B_{\alpha}=G_{\alpha\beta}B^{\beta}\,,\qquad 
G_{\alpha\beta}=B_{\alpha}\circ B_{\beta}\ .
\end{equation}
The `$\circ$' product is defined in \eqref{eq:helicity-contraction-rule} and \eqref{eq:helicitysum-fermion}.

For the form factor with three gluons, the gauge invariant basis has 4 elements and we can choose the basis as 
\begin{equation}
B_1= A_1 C_{23}\,,\qquad B_2= A_2 C_{31} \,,\qquad B_3= A_3 C_{12} \,,\qquad B_4=A_1A_2A_3 \,,
\end{equation}
in which $A_{i}$ and $C_{ij}$ are defined by
\begin{equation}
A_{i} = \frac{\varepsilon_i \cdot p_j }{p_i\cdot p_j} -\frac{\varepsilon_i\cdot p_k}{p_i\cdot p_k} \,, \qquad
C_{ij} = \varepsilon_i\cdot \varepsilon_j -\frac{(p_i\cdot \varepsilon_j)(p_j\cdot \varepsilon_i)}{ p_i\cdot p_j} \,.
\label{eq:AC-def}
\end{equation}
where the $\{i,j,k\}$ in $A_{i}$ are cyclic permutations of $\{1,2,3\}$. For form factors with two external gluons, there is only one gauge invariant basis $B_0=C_{12}$.

\begin{figure}[tb]
\centering
\includegraphics[scale=0.45]{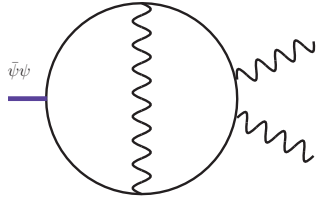}
\caption{A fermion loop containing $\bar{\psi}\psi\in O_{\rm even}$. The gamma trace contains 9 gamma matrices (5 from fermion propagators and 4 from quark-gluon vertices) and thus vanishes.}
\label{fig:odd_operator_fermion_loop}
\end{figure}

Next we consider form factors containing external quarks.
The amplitudes (form factors) with a pair of quark fields contains fermion chains structures like $\bar{u} \pslash{\epsilon}\pslash{p}\cdots u $.  
To define gauge invariant basis, we need to distinguish operators with even and odd numbers of gamma matrices, which will be denoted as $O_{\rm even}$ and $O_{\rm odd}$, respectively. 
For example, the operators $\bar{\psi}\psi$ and $F^{\mu\nu}\bar{\psi}\gamma_{\mu\nu}\psi$ belong to $O_{\rm even}$, while $F^{\mu\nu}D_{\mu}\bar{\psi}\gamma_{\nu}\psi$ belongs to $O_{\rm odd}$. One major difference between $O_{\rm even}$ and $O_{\rm odd}$ is that in a Feynman diagram of a non-chiral massless theory (like Higgs EFT), if $O_{\rm even}$ appears in a fermion loop, the gamma trace of this fermion loop would contain odd number of gamma matrices, thus vanish. An example is shown in Figure \ref{fig:odd_operator_fermion_loop}. 
By contrast, a Feynman diagram with $O_{\rm odd}$ in the fermion loop does not vanish. The scattering amplitudes, and consequently the gauge invariant bases, of $O_{\rm even}$ or $O_{\rm odd}$ contain product of even or odd number of gamma matrices, respectively.

To be more concrete, let us start with the gauge invariant basis for the $H\rightarrow q\bar{q}$ amplitude. In this case, there is only one single element $B_1=\bar{u}(p_2)u(p_1)$ for the basis. Consequently, only form factors of $O_{\rm even}$ contribute and all $O_{\rm odd}\rightarrow q\bar{q}$ type amplitudes must vanish, since the gauge invariant basis for the latter must contain odd number of gamma matrices in the product. The gauge invariant basis for $H\rightarrow q\bar{q}$ can be summarized as:
\begin{equation}
B^{\rm odd}=\{\} \,, \qquad B^{\rm even}=\{\bar{u}(p_2)u(p_1)\} \,.
\end{equation}

For amplitudes $H\rightarrow q\bar{q}g$, the gauge invariant bases contain 4 types of contractions: 
\begin{equation}
\bar{u}(p_2)u(p_1) , \ \bar{u}(p_2)\pslash{p}_3u(p_1), \ \bar{u}(p_2)\pslash{\epsilon}_3u(p_1), \ \bar{u}(p_2)\pslash{\epsilon}_3\pslash{p}_3u(p_1) \,,
\end{equation} 
in which two of them contain odd/even number of gamma matrices. The gauge invariance basis can be constructed as:
\begin{align}
B^{\rm odd} & = \left\{ \bar{u}(p_2)\pslash{p}_3u(p_1)A_3,\ \ \bar{u}(p_2)\pslash{\epsilon}_3u(p_1)(p_1\cdot p_3)-\bar{u}(p_2)\pslash{p}_3u(p_1)(p_1\cdot \epsilon_3) \right\} , \\
B^{\rm even} & = \left\{ \bar{u}(p_2)u(p_1)A_3, \ \ \ \ \ \,  \bar{u}(p_2)\pslash{\epsilon}_3\pslash{p}_3u(p_1)(p_1\cdot p_3) \right\} .
\end{align}

After expanding a form factor in the gauge invariant basis as in \eqref{eq:FprojectedinBandf}, the helicity information is contained in the basis $B_\alpha$, and the coefficients $f^{\alpha}_n$ contain only scalar product of loop and external momenta, which can be reduced directly using IBP. 
Comparing with other tensor reduction methods like the PV reduction, the gauge invariant basis method produces integrals with less numerator power, and the coefficients of the integrals are also more compact and do not contain Gram determinants.

\subsection{Details of the unitarity-IBP construction}
\label{3.3}

In this subsection we demonstrate the strategy described above using explicit examples.
Let us first point out an important difference between planar form factors and planar scattering amplitudes: the planar-color form factors can contain integrals whose topologies are non-planar.  This is because the operator (or the Higgs particle) is a color singlet, thus the presence of the operator does not alter the structure of the color diagram. So the diagram contributes to the leading $N_c$ order even if the operator appears in the middle of the diagram.
Since the operator carries a non-zero momentum $q$, the planar-color diagram for form factors may correspond to a non-planar integral.


\subsubsection*{Two-loop two-gluon form factor}

\begin{figure}[tb]
\centering
\includegraphics[scale=0.4]{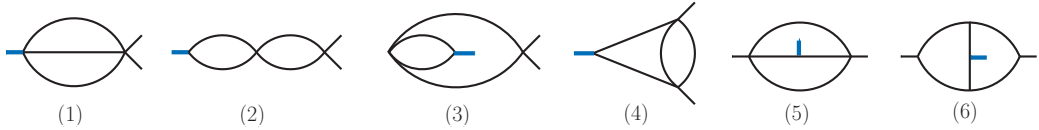}
\caption{The master integrals of the 2-loop 2-point form factor.}
\label{fig:FF2g2loopMIs}
\end{figure}

\begin{figure}[tb]
\centering
\includegraphics[scale=0.4]{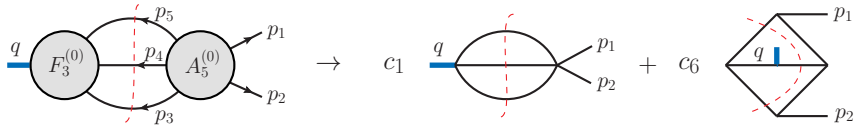}
\caption{The triple cut for a two-loop form factor of ${\rm tr}(F^2)$ with two external particles.}
\label{fig:triplecutExample}
\end{figure}

As a simple example, we consider the two-point two-loop form factor ${\cal O}_0\rightarrow 2g$ in the pure gluon sector.
The complete set of master integrals are given in Figure \ref{fig:FF2g2loopMIs}. One may note that two pairs of master integrals, (2) and (3), and (4) and (5) in Figure \ref{fig:FF2g2loopMIs},  are equivalent up to some relabellings of loop momenta. 
In our unitarity computation, we need to distinguish them, because as discussed at the beginning of this subsection, they correspond to different planar-cut diagram contributions.
Practically, the relabeling of loop momenta is not consistent with cut conditions, since the cut momenta are fixed in a given unitary cut. By choosing not to relabel loop momenta during IBP reduction, the above pairs of integrals will appear as distinguishable master integrals.

Below we demonstrate the unitarity-IBP method by considering the triple-cut shown in the l.h.s. of Figure \ref{fig:triplecutExample}. 
The two-loop cut form factor is given as the product of a color-ordered three-gluon tree form factor and a color-ordered five-gluon tree amplitudes:
\begin{align}
&\sum_{\rm helicities \ of\ \varepsilon_{3,4,5}}F_{3}(-5,-4,-3) A_5(1,2,3,4,5)  \,.
\end{align}
The polarization vectors $\varepsilon_{3,4,5}$ of the cut gluons can be summed using the contraction rule in \eqref{eq:helicity-contraction-rule},
then the polarization vectors $\varepsilon_{1,2}$ of the external gluons can be contracted with the gauge invariant basis, which contains a single element  $B_0=C_{12}$ in \eqref{eq:AC-def}. Thus we have
\begin{align}
\sum_{\rm helicities}F_{3} \, A_5 = \big[ (\varepsilon_1\cdot \varepsilon_2)(p_1\cdot p_2) -(p_1\cdot \varepsilon_2)(p_2\cdot \varepsilon_1) \big] \, f_0(\{s_{ij}, D\})  \,.
\end{align}
The scalar function $f_0$ is a function that is rational in $s_{ij}$ and polynomial in the dimension parameter $D$, and thus it can be directly reduced using IBP reduction with e.g.~public codes \cite{Smirnov:2014hma, Lee:2013mka, vonManteuffel:2012np, Maierhoefer:2017hyi}. As shown in the r.h.s.~of Figure \ref{fig:triplecutExample}, only two master integrals (1) and (6) in Figure \ref{fig:FF2g2loopMIs} enter in this cut. This cut allows us to compute their coefficients $\{c_1, c_6\}$ as
\begin{align}
c_1 =& \  12 D-\frac{1175}{6 (D-4)}-\frac{1}{D-3}+\frac{48}{D-2}+\frac{58}{9 (D-1)}+\frac{525}{32 (2 D-7)}+\frac{107}{288 (2 D-5)} \nonumber\\
& \ -\frac{694}{3 (D-4)^2}-\frac{16}{(D-2)^2}-\frac{96}{(D-4)^3}-\frac{1955}{16} \,, \\
c_6 = & \  \frac{3 (D-3) (3 D-8)}{4 (2 D-7) (2 D-5)} \,,
\end{align}
which are consistent with the known result (see e.g. \cite{Gehrmann:2010ue}).

\begin{figure}[tb]
\centering
\includegraphics[scale=0.4]{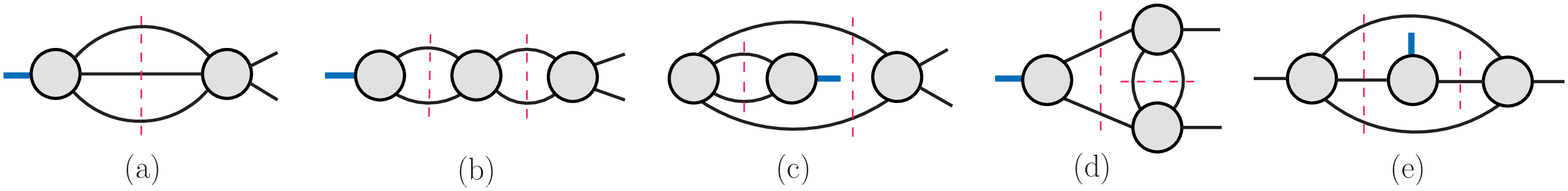}
\caption{The cuts needed in the 2-loop 2-point form factor calculation.}
\label{fig:FF2g2loopAllcuts}
\end{figure}

To determine the coefficients of other master integrals, four other cuts can be used as shown in Figure \ref{fig:FF2g2loopAllcuts}. More explicitly:
cut-(b) for $\{c_2\}$, cut-(c) for $\{c_3\}$, cut-(d) for $\{c_4\}$ and cut-(e) for $\{c_5, c_6\}$. Note that $c_6$ appears in both cut-(a) and (e), and these two cuts provide a non-trivial consistency check.

The full form factor ${\cal F}_{{\cal O}_0}^{(2)}$ can be written as
\begin{equation}
{\cal F}_{{\cal O}_0}^{(2)}(p_1,p_2) = \bigg( \sum_{i=1}^4 c_i M_i + {1\over2}\sum_{i=5,6} c_i M_i \bigg) + \textrm{perms}(p_1, p_2) \,,
\end{equation}
where $M_i$ correspond to the master integrals with label $(i)$ in Figure \ref{fig:FF2g2loopMIs}. Note that the permutation of two external gluons does not alter the integrals (5) and (6), so for these two master integrals a factor $\frac{1}{2}$ is added to avoid double counting.

\subsubsection*{Two-loop three-gluon form factor}

\begin{figure}[tb]
\centering
\includegraphics[scale=0.35]{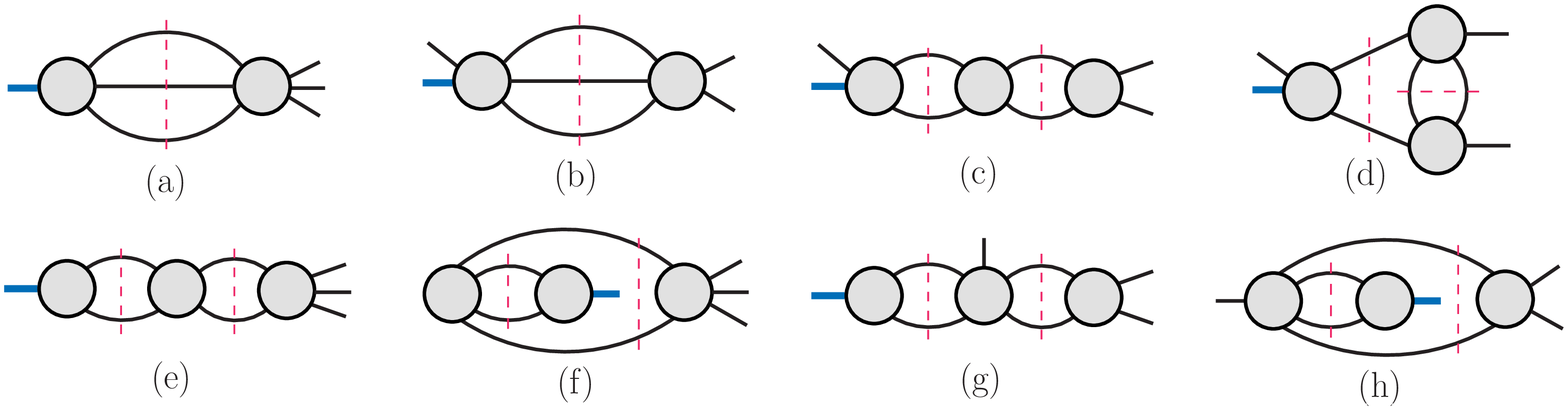}
\caption{The cuts needed in the 2-loop 3-point form factor calculation.}
\label{fig:FF3g2loopAllcuts}
\end{figure}

\begin{figure}[tb]
\centering
\includegraphics[scale=0.38]{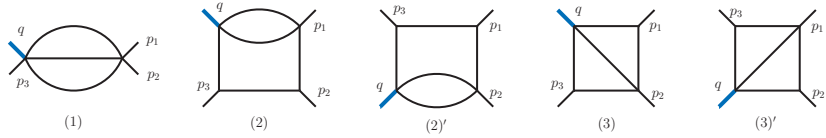}
\caption{Master integrals of ${\cal F}_{{\cal O}_1}^{(2)}$ captured by the $s_{12}$ triple cut (b) in Figure \ref{fig:FF3g2loopAllcuts}.}
\label{fig:s12triplecut}
\end{figure}

\begin{figure}[tb]
\centering
\includegraphics[scale=0.38]{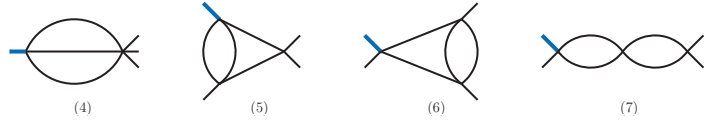}
\caption{Master integrals of ${\cal F}_{{\cal O}_1}^{(2)}$ that are \emph{not} captured by the triple cut (b) in Figure \ref{fig:FF3g2loopAllcuts}.}
\label{fig:nots12triplecut}
\end{figure}

The set of cuts which is sufficient for the computation of the three-point two-loop form factors are given in Figure \ref{fig:FF3g2loopAllcuts}. All these cuts are required for the form factors of length-2 operators, while for length-3 operators only the four cuts in the first row are needed.
Consider the two-loop three-gluon form factor of length-3 operator $\mathcal{O}_1$ as an example.
${\cal F}_{{\cal O}_1}^{(2)}$ contains seven master integrals up to permutations of external legs, as show in Figure \ref{fig:s12triplecut} and Figure \ref{fig:nots12triplecut}. Each cut fixes the coefficients of a subset of these master integrals.  For example, triple cut (b) of Figure \ref{fig:FF3g2loopAllcuts} in $s_{12}$-channel determines the coefficients of five master integrals in Figure \ref{fig:s12triplecut}, and the coefficients of $(2)'$ (or $(3)'$) are related to that of (2) (or (3))  by flipping symmetry $p_1 \leftrightarrow p_2$. If a master integral appears in the results of several different cuts, its coefficient in these cuts must be the same, which provides consistency check for the computation.

\begin{figure}[tb]
\centering
\includegraphics[scale=0.26]{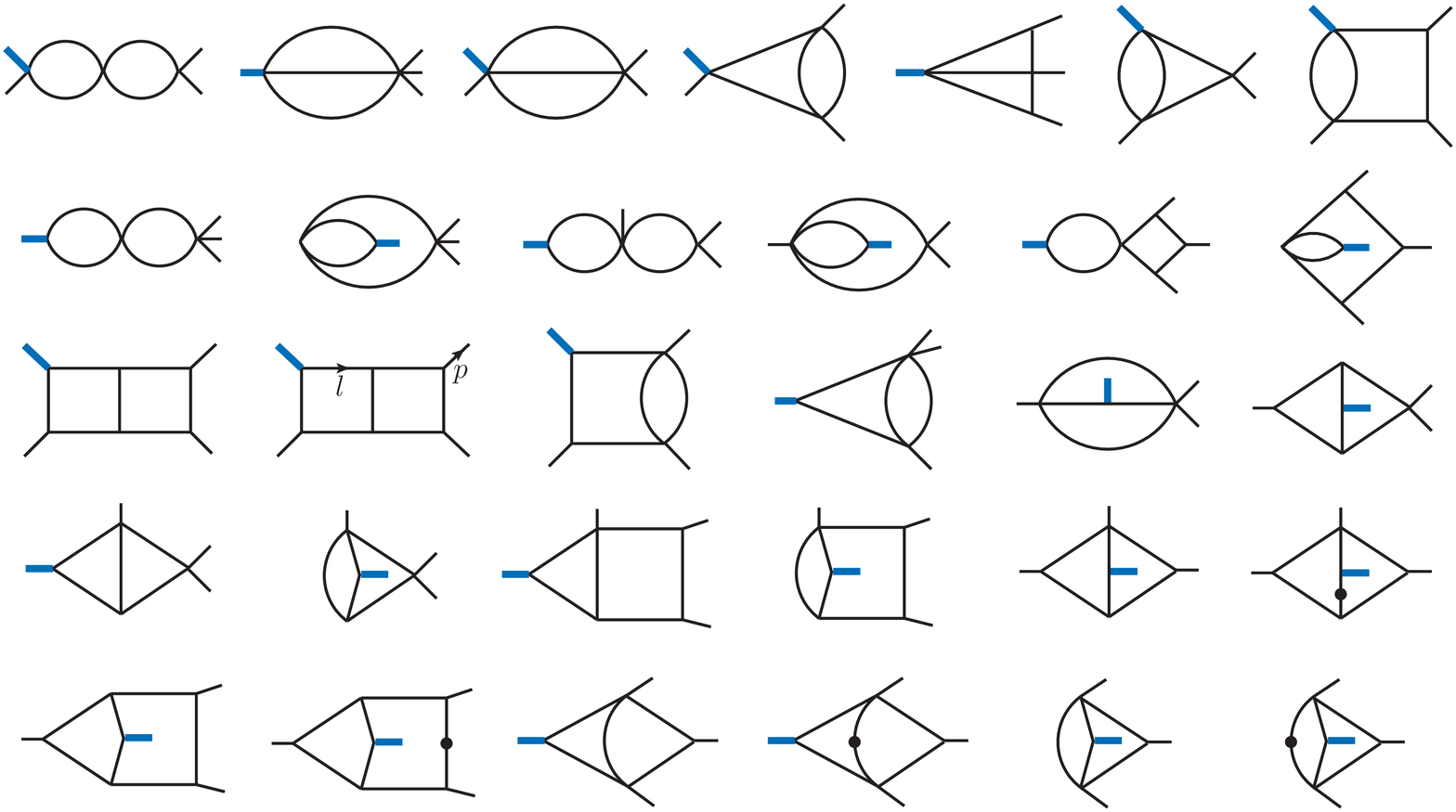}
\caption{The full set of master integrals of the two-loop 3-point form factor. The line with a dot represents a double propagator.}
\label{fig:FF_3g_2loop_MIs}
\end{figure}

For the Higgs to three-parton amplitudes considered in this paper, the  full set of master integrals are shown in Figure \ref{fig:FF_3g_2loop_MIs}. 
They have been obtained in terms of 2d harmonic polylogarithms \cite{Gehrmann:2000zt,Gehrmann:2001ck}. Using these expressions we can obtain the analytic bare form factors.

\section{Color decomposition of fermion cuts}
\label{sec:color-decomp}

In the pure gluon sector, planar cuts are enough to construct the form factors. 
The cut form factor can be decomposed as products of planar tree form factors or amplitudes.
However, such a decomposition is not obvious in the presence of quark loops.
In this section we show that, in the case of Higgs to 3-gluon amplitudes, by making connection between the fundamental and adjoint fermions, a nice color decomposition is still possible, such that the full 2-loop integrand can be constructed using planar cuts.

In our notation, gluons carry an adjoint color index $a=1,2,\ldots,N_c^2-1$, 
and quarks and antiquarks carry an $N_c$ or $\overline{N}_c$ index,
$i, \bar\jmath=1,\ldots,N_c$. 
We will use the group algebras
\begin{equation}
\Tr(T^aT^b)=t_F\delta^{ab}, \qquad  (T^aT^a)_i^{~\bar\jmath} =C_F \delta_i^{~\bar\jmath} , \qquad  (T^aT^bT^a)_i^{~\bar\jmath}= \Big( C_F-\frac{C_A}{2} \Big) (T^b)_i^{~\bar\jmath} \,.
\end{equation}
We denote $\mathrm{f}_x$ for the flavor index of quarks and the contraction is given by $\delta_{\mathrm{f}_x \mathrm{f}_x}=n_f$.

\subsection{Color decomposition of tree amplitudes}

As far as the color factors are concerned, we do not need to discriminate $n$-point amplitudes and $n$-point form factors. Since the Higgs field is a color singlet, we can remove it from the form factor color graph, what is left is the color graph of a scattering amplitude. For example, the $H\rightarrow 3g$ tree form factor has the color factor $f^{abc}$, which is the same as the color factor of 3-gluon tree amplitude.

We use the following color decomposition of $n$-gluon tree amplitudes:
\begin{equation}
\mathcal{A}(1_g,2_g,\cdots ,n_g)= 
\sum_{\sigma\in S_{n-2}}A\Bigl((n-1)\sigma_1\sigma_2\cdots \sigma_{n-2}n\Bigr)f^{a_{n-1}a_{\sigma_1}\cdots a_{\sigma_{n-2}}a_n} \,.
\end{equation}
Here $\mathcal{A}$ ($\mathcal{F}$) denotes the amplitudes (form factors) with full color factors, while $A$ ($F$) denotes the color-tripped planar amplitudes (form factors).

We also need tree amplitudes and form factors with quark pairs. For tree amplitudes with one quark pair and $(n-2)$ gluons, a similar color decomposition is
\begin{equation}
\mathcal{A}(1_g,2_g,3_g,\cdots ,(n-1)_q, n_{\bar{q}})=\sum_{\sigma\in S_{n-2}}A\Big( (n-1)_q \sigma(1)\cdots \sigma(n-2) n_{\bar{q}}\Big )\big(T^{a_{\sigma_1}}\cdots T^{a_{\sigma_{n-2}}}\big)_{i_{n-1}}^{\ \ \  \bar\imath_{n}}\delta_{\mathrm{f}_1 \mathrm{f}_n} \,.
\end{equation}
The color decomposition for the 4-quark tree amplitude is
\begin{equation}
\mathcal{A}(1_q, 2_q, 3_{\bar{q}}, 4_{\bar{q}})=A(1342)(T^a)_{i_1}^{~\bar\imath_3}(T^a)_{i_2}^{~\bar\imath_4}\delta_{\mathrm{f}_1 \mathrm{f}_3}
\delta_{\mathrm{f}_2 \mathrm{f}_4}
+A(1432)(T^a)_{i_1}^{~\bar\imath_4}(T^a)_{i_2}^{~\bar\imath_3}\delta_{\mathrm{f}_1 \mathrm{f}_4}
\delta_{\mathrm{f}_2 \mathrm{f}_3} \,.
\end{equation}

\subsection{The $s_{12}$ two double-cut}

\begin{figure}[tb]
\centering
\includegraphics[scale=0.57]{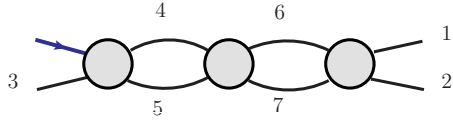}
\caption{The $s_{12}$ two double-cut.}
\label{fig:s12doublebubblecut}
\end{figure}

The $s_{12}$ two double-cut, as show in Figure \ref{fig:s12doublebubblecut}, corresponds to the product of a 3-point form factor and two 4-point amplitudes:
$ \sum \mathcal{F}(345) \, \mathcal{A}(7654) \, \mathcal{A}(1267) $.

First we consider the case when all cut lines $(4,5,6,7)$ are all fermions. The cut integrand is the product of the following three tree amplitudes:
\begin{align}
\mathcal{F}(3_g, 4_{\bar{q}}, 5_q)= & \ F(3_g,4_{\bar{q}},5_q)(T^{a_3})_{i_5}^{~\bar\imath_4}\delta_{\mathrm{f}_5 \mathrm{f}_4} \,, \\
\mathcal{A}(1_g, 2_g, 6_q, 7_{\bar{q}})= &\  \Big[A(1_g,2_g, 7_{\bar{q}}, 6_q)(T^{a_1}T^{a_2})_{i_6}^{~\bar\imath_7}+A(2_g, 1_g, 7_{\bar{q}}, 6_q)(T^{a_2}T^{a_1})_{i_6}^{~\bar\imath_7}\Big]\delta_{\mathrm{f}_6 \mathrm{f}_7} \,,
\\
\mathcal{A}(7_q, 6_{\bar{q}}, 5_{\bar{q}}, 4_q)= & \ A(5_{\bar{q}}, 4_q, 7_q, 6_{\bar{q}})(T^{a})_{i_4}^{~\bar\imath_5}(T^{a})_{i_7}^{~\bar\imath_6}\delta_{\mathrm{f}_4 \mathrm{f}_5} \delta_{\mathrm{f}_7 \mathrm{f}_6}
+(4\leftrightarrow 7) \,.
\end{align}

After contracting the color and flavor indices, the cut integrand can be reduced to
\begin{align}
\mathcal{F}\Bigr|^{qqqq}_{\rm cut}=&\ F(3_g,4_{\bar{q}}, 5_q)
\bigg[A(5_{\bar{q}}, 4_q, 7_q, 6_{\bar{q}}) \, n_f^2 \,t_F
+A(5_{\bar{q}}, 7_q, 4_q, 6_{\bar{q}}) \Big( C_F-\frac{C_A}{2} \Big) n_f \bigg]
\nonumber\\
& \ \times\Bigl[A(1_g, 2_g, 7_{\bar{q}}, 6_q)\Tr(T^{a_1}T^{a_2}T^{a_3})
+A(2_g,1_g,7_{\bar{q}}, 6_q)\Tr(T^{a_1}T^{a_3}T^{a_2})\Bigr] \,,
\label{fcutqqqq}
\end{align}
where the superscript $qqqq$ indicates that the four cut legs are all quarks.
We can see that the product of tree amplitudes apparently do not have planar structure.
Four different color structures appear in this configuration, and we rewrite \eqref{fcutqqqq} as
\begin{align}
\mathcal{F}\Bigr|^{qqqq}_{\rm cut}= & \ \Bigl[c_1 n_f^2 \,t_F+c_2 n_f (C_F-\frac{C_A}{2})\Bigr]\Tr(T^{a_1}T^{a_2}T^{a_3}) \nonumber\\
& \ +\Bigl[c_3 n_f^2 \, t_F+c_4 n_f (C_F-\frac{C_A}{2})\Bigr]\Tr(T^{a_1}T^{a_3}T^{a_2}) \,,
\label{fcutqqqq2}
\end{align}
where the kinematic parts are absorbed in the $c_i$ factors.

It is important to notice that our discussion so far applies to general representation of quarks. In the case that quarks are in adjoint representation, it is clear that the cut integrand is proportional to $C_A^2 \, f^{a_1a_2a_3}$ and can be written as
\begin{equation}
\label{fcutqqqq3}
\mathcal{F}_{\rm adj}\Bigr|^{qqqq}_{\rm cut}=(X_1 \, n_f^2+X_2 \, n_f) \, C_A^2 \, f^{a_1a_2a_3} \,,
\end{equation}
where the kinematic parts $X_i$ can be computed using planar unitarity cuts and correspond to the coefficients of $n_f^2$ and $n_f$ in the planar cut integrand, respectively.

In order to match \eqref{fcutqqqq2} with \eqref{fcutqqqq3} when taking fermions to be adjoint, we must have $c_3=-c_1$ and $c_4=-c_2$, and \eqref{fcutqqqq2} can be reduced to
\begin{equation}
\label{fcutqqqq3-2}
\mathcal{F}\Bigr|^{qqqq}_{\rm cut}=i\Bigl[c_1 n_f^2 \,t_F+c_2 n_f (C_F-\frac{C_A}{2})\Bigr] t_F\, f^{a_1a_2a_3} \,.
\end{equation}
Furthermore, in the adjoint fermion case, the two color factors above reduce to:
\begin{equation}
t_F\rightarrow C_A \,, \qquad C_F-\frac{C_A}{2}\rightarrow\frac{ C_A}{2} \,.
\end{equation} 
By matching the $n_f^2$ and $n_f$ terms, the result in a generic representation can be written as
\begin{align}
\mathcal{F}\Bigr|^{qqqq}_{\rm cut} &= \bigg[ X_1 \, n_f^2 \, t_F^2+X_2 \, n_f \, t_F \Big(2C_A-{C_F} \Big)\bigg] f^{a_1a_2a_3} \,.
\end{align}
So the cut integrand can be obtained using planar unitarity cut in the adjoint case. To be more explicit: first, one compute the cut integrand in the adjoint representation using planar unitarity cut, then replace $C_A^2$ by $t_F^2$ in the coefficient of $n_f^2$, and replace $C_A^2$ by $t_f\Big(C_A-\frac{C_F}{2} \Big)$ in the coefficient of $n_f$.

In the case that $(4,5)$ are gluons, and $(6,7)$ are fermions, one obtains the following color decomposition:
\begin{align}
\mathcal{F}\Bigr|^{ggqq}_{\rm cut}=&\ F(3_g,4_g,5_g)\frac{iC_A}{2}n_f \, t_F\Bigl[A(4_g,5_g,6_{\bar{q}},7_q)-A(5_g,4_g,6_{\bar{q}},7_q)\Bigr] \nonumber\\
&\ \times \Bigl[A(1_g, 2_g, 7_{\bar{q}}, 6_q)\Tr(T^{a_1}T^{a_2}T^{a_3})
+A(2_g, 1_g, 7_{\bar{q}}, 6_q)\Tr(T^{a_2}T^{a_1}T^{a_3})\Bigr]\, ,
\end{align}
which has two color structures, $C_At_F\Tr(T^{a_1}T^{a_2}T^{a_3})$ and $C_At_F\Tr(T^{a_1}T^{a_3}T^{a_2})$. The same structure happens in the case $(4,5)$ are fermions, and $(6,7)$ are gluons. 
By similar analysis as the previous example, if we denote the cut amplitude in adjoint representation as
\begin{equation}
\mathcal{F}_{\rm adj}\Bigr|^{ggqq}_{\rm cut}=X_3 \, n_f \, C_A^2 \, f^{a_1a_2a_3} \,, \qquad
\mathcal{F}_{\rm adj}\Bigr|^{qqgg}_{\rm cut}=X_4 \, n_f \, C_A^2 \, f^{a_1a_2a_3} \,,
\end{equation}
the cut amplitude in generic representation can be written as
\begin{equation}
\mathcal{F}\Bigr|^{ggqq}_{\rm cut}=X_3 \, n_f \, t_F \, C_A \, f^{a_1a_2a_3} \,, \qquad 
\mathcal{F}\Bigr|^{qqgg}_{\rm cut}=X_4 \, n_f \, t_F \, C_A \, f^{a_1a_2a_3} \,.
\end{equation}
Here again, $X_3$ and $X_4$ can be extracted form the cut integrand in adjoint representation.

The above discussion means that the planar cut is suffice to determine the $s_{12}$ double 2-cut integrand in the generic representations.
The only difference is that different color factors should be assigned to different terms in the cut integrand. 
All these color factors should be reduced to $C_A^2$ in the adjoint representation.

\subsection{The $s_{12}$ triple-cut}

\begin{figure}[tb]
\centering
\includegraphics[scale=0.5]{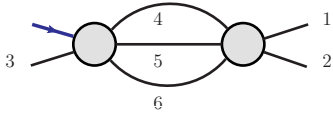}
\caption{The $s_{12}$ triple-cut}
\label{fig:s12sunrisecut}
\end{figure}
\begin{figure}[tb]
\centering
\includegraphics[scale=0.45]{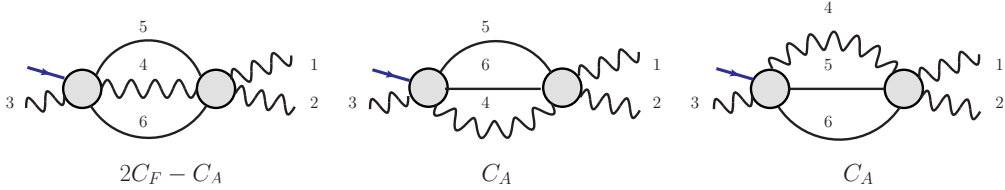}
\caption{Different ordering of fields corresponds to different color factors.}
\label{fig:s12sunrisecolor}
\end{figure}

The $s_{12}$ triple-cut corresponds to the product of a 4-point form factor and a 5-point amplitudes, $\mathcal{F}(3456)\mathcal{A}(12456)$, as show in Figure \ref{fig:s12sunrisecut}. If the internal states are all gluons, the color factor is simply $C_A^2$.
Now consider the case $(4,5,6)=(g, q, \bar{q})$. The tree amplitudes are
\begin{align}
\mathcal{F}(3_g, 4_g, 5_{\bar{q}}, 6_q) & =F(3_g, 4_g, 5_{\bar{q}}, 6_q)(T^{a_3}T^{a_4})_{i_6}^{~\bar\imath_5}\delta_{\mathrm{f}_6  \mathrm{f}_5}+(3\leftrightarrow 4)\, , \nonumber\\
\mathcal{A}(1_g, 2_g, 4_g, 5_q, 6_{\bar{q}}) & =
A(1_g, 2_g, 4_g, 6_{\bar{q}}, 5_q)(T^{a_1}T^{a_2}T^{a_4})_{i_5}^{~\bar\imath_6}\delta_{\mathrm{f}_5 \mathrm{f}_6}
+\text{permutations of (124)} \,.
\label{s12sunrise1}
\end{align}
Contracting the color and flavor indices, and using  the $U(1)$ decoupling relation
\begin{equation}
A(14265)=-A(12465)-A(12645)-A(12654)\ ,
\end{equation}
the cut amplitude can be rewritten as
\begin{align}
\mathcal{F}\Bigr|^{gqq}_{\rm cut}=\frac{n_f}{2}\Tr(T^{a_1}T^{a_2}T^{a_3})\Bigl[ &
(2C_F-C_A)F(3546)A(12645)+C_AF(4356)A(12465) \nonumber\\
 & + C_AF(3456)A(41265)\Bigr]+(1\leftrightarrow 2) \,. 
\label{s12sunrise3}
\end{align}
The three terms in the bracket in \eqref{s12sunrise3} take obviously the planar-cut form and correspond to the three different internal-state configurations in Figure \ref{fig:s12sunrisecolor}, respectively. If the gluon line appears in the middle of the diagram, the color factor is $2C_F-C_A$, otherwise it is $C_A$. The same pattern also appears in other cuts.

\subsection{Other cuts}

The color structures of the other cuts can be computed in a similar way. It turns out that for every cut, a planar color decomposition is possible. We summarize all cases as follows (where (a)-(h) correspond to labels in Figure \ref{fig:FF3g2loopAllcuts}):
\begin{enumerate}
\item $s_{123}$ two double-cut (e)-(f): both planar and non-planar cases have factor $C_F$.

\item $s_{12}$ two double-cut (c): 4-fermion cut has color factors $n_f (2C_F-C_A)$ and $n_f^2t_F$, other channels have factor $C_A$.

\item $s_{123}$ triple-cut (a), $s_{12}$ triple-cut (b), $s_{12}$ triangle-bubble-cut (d): diagrams in which gluon appears in the middle have factor $2C_F-C_A$, other channels have factor $C_A$.

\item $s_{12}-s_{123}$ two double-cut (g)-(h): the nonplanar case has factor $2C_F-C_A$, the planar case has factor $C_A$.

\end{enumerate}
These allow us to compute full amplitudes using only planar cuts.

\section{Results}
\label{sec:results}

In this section we perform UV renormalization and IR subtraction for form factors and obtain compact analytic forms.
The two-loop bare form factors contain UV and IR divergences, which were discussed in Section \ref{sec:divergence-structure}. The $1/\epsilon^m, m=4,3,2$ pole terms must cancel with the universal IR divergences and the 1 loop UV divergences, which offer non-trivial self-consistency checks of the results. 
The cancellation of $1/\epsilon$ pole terms then determines the two-loop anomalous dimension of the operator. 

As an important check of the computation, we have reproduced known results including the non-trivial two-loop amplitudes of Higgs to three partons with the operator ${\rm tr}(F^2)$ \cite{Gehrmann:2011aa}. 
As a further check, we recall that the form factors should satisfy the linear relation \eqref{eq:O1-linear-relation}.
We compute form factors of different operators independently.
We explicitly check that, already at the level of IBP master integrals, the results satisfy exactly this linear relation. 
We would like to emphasize the computation of form factors of ${\rm tr}(D^2F^2)$ is more involving than the known result of ${\rm tr}(F^2)$ due to the extra derivatives in the operator, and our method can be efficiently applied to such case as well as operators with higher dimensions.

A word about notation: for form factors with three partons, it is enough to consider three configurations given in Table \ref{table:FF-notation}. The subscripts $\alpha,\beta,\gamma$ denote different external states, similar to that in \cite{Gehrmann:2011aa}. 
We also introduce dimensionless variables:
\begin{align}
u = {s_{12} \over s_{123} } \,, \quad v = {s_{23} \over s_{123}} \,, \quad w = {s_{13} \over s_{123}} \,, \qquad \textrm{where} \ \ s_{123} = q^2 = m_H^2 \,.
\end{align}

\begin{table}[t]
\caption{Notation of form factors with three partons, where $\pm$ indicates positive or negative helicity gluons.}
\label{table:FF-notation}
\begin{center}
\begin{tabular}{c|c|c|c}
\hline
external particles & $(1^-, 2^-, 3^-)$ & $(1^-, 2^-, 3^+)$ & $(1^q, 2^{\overline{q}}, 3^-)$ \\ \hline 
form factors & ${\cal F}^{(l)}_{{\cal O}_i, \alpha}$ & ${\cal F}^{(l)}_{{\cal O}_i, \beta}$ & ${\cal F}^{(l)}_{{\cal O}_i, \gamma}$ \\ 
\hline
\end{tabular}
\end{center}
\end{table}

\subsection{Tree-level results}

We first recall the tree-level form factors for the dimension-4 operator ${\rm tr}(F^2)$ \cite{Dixon:2004za}:
\begin{align}
& {\cal F}_{{\cal O}_0,\alpha}^{(0)} = {s_{123}^2 \over [1 2] [2 3] [3 1]}\,, \qquad  {\cal F}_{{\cal O}_0,\beta}^{(0)} =  { \langle 1 2\rangle^4 \over \langle 1 2\rangle \langle 2 3 \rangle \langle 3 1 \rangle} \,, \qquad {\cal F}_{{\cal O}_0,\gamma}^{(0)} = { \langle 2 3\rangle^2 \over \langle 1 2\rangle } \,.
\end{align}
Since the operators satisfy the linear relation \eqref{eq:ope-linear-relation},
it is convenient to introduce a dimension-6 operator $\hat{\cal O}_2$ as
\begin{equation}
\hat{\cal O}_{2} \equiv \partial^2{\cal O}_{0} \,.
\end{equation}
The form factor of $\hat{\cal O}_2$ is the same as ${\cal O}_{0}$ up to an overall factor $s_{123}$:
\begin{equation}
\label{eq:FO2viaFO0}
{\cal F}_{\hat{\cal O}_{2}} = s_{123} \, {\cal F}_{{\cal O}_{0}} \,.
\end{equation}

For convenience, we normalize all form factors of dimension-6 operators by dividing the tree form factor of $\hat{\cal O}_2$ and introduce the  `dimensionless' form factors $r^{(\ell)}_{\hat{\cal O}_I}$ as
\begin{equation}
r^{(\ell)}_{\hat{\cal O}_I} := {{\cal F}_{b, \hat{\cal O}_I}^{(\ell)}} / {{\cal F}_{\hat{\cal O}_2}^{(0)} } \,,
\label{eq:r-ratioFF-def}
\end{equation}
%
\begin{table}[t]
\caption{Normalized tree-level form factors $r^{(0)}_{\hat{\cal O}_I} = {{\cal F}_{\hat{\cal O}_I}^{(0)}} / {{\cal F}_{\hat{\cal O}_2}^{(0)} }$.}
\label{table:tree-results}
\begin{center}
\begin{tabular}{c|c|c|c}
\hline
$r^{(0)}_{\hat{\cal O}}$ & $\alpha$ & $\beta$ & $\gamma$ \\ \hline 
${\hat{\cal O}_1}$ & $u\, v\, w$ & $0$ & $0$  \\ \hline
${\hat{\cal O}_2}$ & $1$ & $1$ & $1$ \\ \hline
${\hat{\cal O}_3}$ & $0$ & $0$ & $0$  \\ \hline
${\hat{\cal O}_4}$ & $0$ & $0$ & $u$  \\ 
\hline
\end{tabular}
\end{center}
\end{table}
%
The ratio tree-level form factors are given as (also summarized in Table \ref{table:tree-results}):
\begin{align}
& r_{\hat{\cal O}_1,\alpha}^{(0)} 
= u\, v \, w\,, \qquad  \qquad \qquad r_{\hat{\cal O}_1,\beta}^{(0)} = r_{\hat{\cal O}_1,\gamma}^{(0)} = 0 \,, \\
& r_{\hat{\cal O}_3,\alpha}^{(0)} =  r_{\hat{\cal O}_3,\beta}^{(0)} = r_{\hat{\cal O}_3,\gamma}^{(0)} = 0 \,, \\
& r_{\hat{\cal O}_4,\alpha}^{(0)} =  r_{\hat{\cal O}_4,\beta}^{(0)} = 0 \,, \qquad \qquad r_{\hat{\cal O}_4,\gamma}^{(0)} = u \,.
\end{align}
Note that, we have normalized the operators $\{ \hat{\cal O}_1, \hat{\cal O}_3, \hat{\cal O}_4 \}$ properly, such that the 3-point tree form factors all have the unit constant.
From now on, we will take $\hat{\cal O}_I$:
\begin{equation}
\hat{\cal O}_I = \{ \hat{\cal O}_1, \hat{\cal O}_2,  \hat{\cal O}_3, \hat{\cal O}_4 \} \,,
\end{equation}
as the basis of dimension-6 operators.

\subsection{Loop corrections}

Below we consider the form factors of ${\cal O}_0$ and $\hat{\cal O}_1$ in detail. 
Explicit results of other operators are collected in Appendix \ref{app:oneloop} -- \ref{app:remainderO4}.

\subsubsection*{Example 1: ${\cal F}_{{\cal O}_0}(1^-, 2^-, 3^-)$}

We first consider the form factor of ${\cal O}_0$ and three gluons. This result has been obtained in \cite{Gehrmann:2011aa}. 
The main purpose of this discussion is to make contact with the known literature and to set up the notation which will be used for the higher dimension operator cases. Since ${\cal O}_0$ is a length-2 operator, we have $\delta_n = 3 - 2 =1$, as defined in Section \ref{sec:divergence-structure}.

The one-loop bare form factor is:
\begin{equation}
{\cal F}_{b,{\cal O}_0,\alpha}^{(1)} = {\cal F}_{{\cal O}_0,\alpha}^{(0)} \Big( a_1 I_4[1,2,3,q] + b_1 I_2[s_{12}] + c_1 I_2[s_{123}] + \textrm{(cyclic perm.)}  \Big) \,, 
\end{equation}
where $I_4$ and $I_2$ are one-loop box and bubble master integrals, and the master coefficients to all order in $\epsilon$ are
\begin{align}
a_1 = & {u^2 v^2 \over 2 w}\frac{ \epsilon ^2}{(2 \epsilon -1)}-\frac{u v}{2}  \,, \\
b_1 = & -  n_f\frac{v w \epsilon}{(2 \epsilon-3)(\epsilon-1)}+u \epsilon  \left(\frac{v}{w}+\frac{w}{v}+\frac{\epsilon }{1-\epsilon }\right)+\frac{v w \epsilon }{3-2
   \epsilon }+\frac{1}{\epsilon }-2 \,, \\
c_1 = & \frac{1}{3} \epsilon  \left(-\frac{v w}{u}-\frac{u v}{w}-\frac{u w}{v}+3\right)+\frac{1}{\epsilon -1}-\frac{1}{\epsilon }+3 \,.
\end{align}

As discussed in Section \ref{sec:divergence-structure}, the one-loop form factor satisfies the following relation (with $\delta_n=1$): 
\begin{equation}
\label{eq:F1loop_ren}
{\cal F}^{(1)} = S_\epsilon^{-1} {\cal F}_{\rm b}^{(1)} +  \Big( Z^{(1)} - {\beta_0 \over 2\epsilon} \Big) {\cal F}_{\rm b}^{(0)}   = I^{(1)}(\epsilon) {\cal F}^{(0)} + {\cal F}^{(1),{\rm fin}} + {\cal O}(\epsilon) \,.
\end{equation}
Using the bare one-loop form factor and universal IR information, we can extract the one-loop renormalization constant
\begin{equation}
Z_{{\cal O}_0}^{(1)} = - {1\over \epsilon} \Big( {11 C_A \over 3} - {2 n_f \over 3} \Big)  = - {\beta_0 \over \epsilon} \,.
\end{equation}
The one-loop finite remainder can be obtained as
\begin{equation}
{\cal F}_{{\cal O}_0,\alpha}^{(1),{\rm fin}} = {\cal F}_{{\cal O}_0,\alpha}^{(0)} \Big( N_c\, {\cal R}_{{\cal O}_0,\alpha}^{(1),N_c} + n_f \,{\cal R}_{{\cal O}_0,\alpha}^{(1),n_f}  \Big)  \,,
\end{equation}
where
\begin{align}
{\cal R}_{{\cal O}_0,\alpha}^{(1),N_c}  = & \ -2 \text{Li}_2(1-u)-2 \text{Li}_2(1-v)-2 \text{Li}_2(1-w)-\frac{11}{6} \log (u v w)+\frac{u v}{3}-\log (u) \log (v) \nonumber\\
& \ +\frac{u   w}{3}-\log (u) \log (w)+\frac{v w}{3}-\log (v) \log (w)+\frac{\pi ^2}{2}-\frac{11 \log (-q^2)}{2} \,, \\
{\cal R}_{{\cal O}_0,\alpha}^{(1),n_f}  = & \  \frac{1}{3} \log (u v w)-\frac{u v}{3}-\frac{u w}{3}-\frac{v w}{3} + \log (-q^2) \,.
\end{align}

Similarly, the two-loop form factor satisfies the relation:
\begin{align}
{\cal F}^{(2)} & = S_\epsilon^{-2} {\cal F}_{\rm b}^{(2)} + {S_\epsilon^{-1}} \Big[ Z^{(1)} - {3\over2} {\beta_0 \over \epsilon} \Big] {\cal F}_{\rm b}^{(1)}  +  \Big[ Z^{(2)} -  {\beta_0 \over 2\epsilon} Z^{(1)} + {3\over8} {\beta_0^2 \over \epsilon^2} -  {\beta_1 \over 4\epsilon} \Big] {\cal F}_{\rm b}^{(0)} \nonumber\\
& = I^{(2)}(\epsilon) {\cal F}^{(0)} +  I^{(1)}(\epsilon) {\cal F}^{(1)} + {\cal F}^{(2),{\rm fin}} + {\cal O}(\epsilon)  \,.
\label{eq:F2loop_ren}
\end{align}
Evaluating the bare two-loop form factor and using the universal IR information and one-loop results, 
we can extract the two-loop renormalization constant. The $1/\epsilon^2$ part is determined by the one-loop data as in \eqref{eq:Z2loop_eps2}, while the $1/\epsilon$ part is
\begin{equation}
Z_{{\cal O}_0}^{(2)} \big|_{{1\over\epsilon}\textrm{-part.}} = - {1\over \epsilon} \Big( {34 C_A^2 \over 3} - {10 C_A n_f \over 3} - 2 C_F n_f \Big)  = - {\beta_1 \over \epsilon} \,.
\end{equation}
The two-loop finite remainder can be decomposed according to the color factors as
\begin{equation}
r_{{\cal O}_0,\alpha}^{(2),{\rm fin}} = N_c^2\, {\cal R}_{{\cal O}_0,\alpha}^{(2),N_c^2} + N_c\, n_f\, {\cal R}_{{\cal O}_0,\alpha}^{(2),N_c n_f} + {n_f \over N_c}\, {\cal R}_{{\cal O}_0,\alpha}^{(2),n_f/N_c} + n_f^2\, {\cal R}_{{\cal O}_0,\alpha}^{(2),n_f^2}   \,.
\end{equation}
The explicit expressions are given in \cite{Gehrmann:2011aa} (see also \cite{Duhr:2012fh}), which we do not reproduce here.\footnote{In our notation, $r_{{\cal O}_0}^{(2),{\rm fin}}$ corresponds to $\Omega^{(2),finite}$ in  \cite{Gehrmann:2011aa}.}

\subsubsection*{Example 2: ${\cal F}_{\hat{\cal O}_1}(1^-, 2^-, 3^-)$}

Next we consider the form factor of $\hat{\cal O}_1$. Since $\hat{\cal O}_1$ is a length-3 operator, we have $\delta_n = 3 - 3 =0$. 
The one-loop bare form factor is given in terms of bubble integrals:
\begin{equation}
{\cal F}_{b,\hat{\cal O}_1,\alpha}^{(1)} = {\cal F}_{\hat{\cal O}_1,\alpha}^{(0)} {-6 + 10\epsilon - 4 \epsilon^2 - \epsilon^3 \over 2\epsilon (1-\epsilon)(3-2\epsilon) } (I_2[s_{12}]  + I_2[s_{23}] + I_2[s_{13}] ) \,.
\end{equation}
The one-loop form factor satisfies (with $\delta_n=0$)
\begin{equation}
{\cal F}^{(1)} = S_\epsilon^{-1} {\cal F}_{\rm b}^{(1)} +  Z^{(1)} {\cal F}_{\rm b}^{(0)}   = I^{(1)}(\epsilon) {\cal F}^{(0)} + {\cal F}^{(1),{\rm fin}} + {\cal O}(\epsilon) \,,
\end{equation}
from which we extract the one-loop renormalization constant
\begin{equation}
Z_{\hat{\cal O}_1}^{(1)} = {1\over \epsilon} \Big( {C_A\over2}+n_f \Big)  \,.
\end{equation}

At two loops, using  \eqref{eq:F2loopIR} with $\delta_n=0$, the form factor satisfies
\begin{align}
{\cal F}^{(2)} & = S_\epsilon^{-2} {\cal F}_{\rm b}^{(2)} + {S_\epsilon^{-1}} \Big[ Z^{(1)} - {\beta_0 \over \epsilon} \Big] {\cal F}_{\rm b}^{(1)} + Z^{(2)} {\cal F}_{\rm b}^{(0)} \nonumber\\
& = I^{(2)}(\epsilon) {\cal F}^{(0)} +  I^{(1)}(\epsilon) {\cal F}^{(1)} + {\cal F}^{(2),{\rm fin}} + {\cal O}(\epsilon)  \,.
\label{eq:F2loop_ren_x0}
\end{align}
The cancellation of divergences fixes the two-loop renormalization constant. The $1/\epsilon^2$ part is determined by the one-loop data as in \eqref{eq:Z2loop_eps2},
while the $1/\epsilon$ part presents interesting new structure of operator mixing:
\begin{equation}
(Z^{(2)})_{1}^{~J} \times r_{{\hat{\cal O}_J},\alpha}^{(0)} \big|_{{1\over\epsilon}\textrm{-part.}} =  {1\over \epsilon} \Big( {25 N_c^2\over 12} + {5 N_c n_f\over 12} - {3 n_f\over 4 N_c} \Big) r_{\hat{\cal O}_1,\alpha}^{(0)} - {1\over\epsilon} N_c^2 \, r_{\hat{\cal O}_2,\alpha}^{(0)} \,.
\label{eq:ZO1-mixing}
\end{equation}
We can see  that the first term provides a diagonal part of the renormalization constant matrix:
\begin{align}
({Z^{(2)}})_1^{~1} \big|_{{1\over\epsilon}\textrm{-part.}} = & {1\over \epsilon} \Big( {25 N_c^2\over 12} + {5 N_c n_f\over 12} - {3 n_f\over 4 N_c} \Big) \,,
\end{align}
while the second term is due to the mixing with $\hat{\cal O}_2$ which gives an off-diagonal component of the renormalization constant matrix:
\begin{align}
({Z^{(2)}})_1^{~2} = & - {1\over\epsilon} N_c^2 \,.
\end{align}

The two-loop finite remainder can be further simplified using symbol techniques \cite{Goncharov:2010jf}. We decompose it according to the color factors as
\begin{equation}
{\cal F}_{\hat{\cal O}_1,\alpha}^{(2),{\rm fin}} = {\cal F}_{\hat{\cal O}_1}^{(0)}  \Big( N_c^2\, {\cal R}_{\hat{\cal O}_1,\alpha}^{(2),N_c^2}  + N_c\, n_f\, {\cal R}_{\hat{\cal O}_1,\alpha}^{(2),N_c n_f} + n_f^2\, {\cal R}_{\hat{\cal O}_1,\alpha}^{(2),n_f^2} + {\cal R}^{(2)}_{{\hat{\cal O}_1},\alpha;\log (-q^2)}  \Big) \,,
\end{equation}
where the explicit expressions are collected in Appendix \ref{app:remainderO1}.

Form factors of other operators and other external states can be obtained following the same procedure.
Similar operator mixing effects also appear in other form factors, and we summarize  the renormalization matrix in Section \ref{sec:opemixing}.
The one-loop results in master expansion are collected in Appendix \ref{app:oneloop}. The two-loop finite remainders are collected in Appendix \ref{app:remainderO1} -- \ref{app:remainderO4}.

\subsection{Operator mixing}
\label{sec:opemixing}

As shown in \eqref{eq:ZO1-mixing},
the operators in general have operator mixing effects. This is represented by the renormalization constant matrix $Z_I^{~J}$ defined through
\begin{equation}
\hat{\cal O}_I^{\rm Ren} = Z_I^{~J} \hat{\cal O}_J^{\rm Bare} \,.
\end{equation}
We summarize below the renormalization constant matrix for dimension-6 operators at one and two loops.

At one-loop, there is no operator mixing and the renormalization constant matrix is diagonal:
\begin{align}
(Z_{\hat{\cal O}}^{(1)}) ={1\over\epsilon}\begin{pmatrix}
{N_c\over 2} + n_f & 0  & 0 & 0  \\
0 & - \beta_0 & 0 & 0 \\
0 & 0 & (Z^{(1)})_{3}^{~3} & 0  \\
0 & 0 & 0 &  {8 C_F \over 3} + {2n_f \over 3}
\end{pmatrix} \,.
\nonumber
\end{align}

The two-loop renormalization $Z^{(2)}$ contains  $1/\epsilon^2$ pole terms which are determined by the one-loop matrix using \eqref{eq:Z2loop_eps2}.
The simple pole terms are the intrinsic new two-loop contribution, which can be summarized as follows:
\begin{align}\label{matrix}
& (Z_{\hat{\cal O}}^{(2)}) \big|_{{1\over\epsilon}\textrm{-part.}} = \\
&
{1\over\epsilon}\begin{pmatrix}
{25 N_c^2\over 12} + {5 N_c n_f\over 12} - {3 n_f\over 4 N_c} & - N_c^2 & (Z^{(2)})_{1}^{~3} & {5\over9} + {5 N_c^2 \over 12}  \\
0 & - \beta_1 & 0 & 0 \\
0 & 0 & (Z^{(2)})_{3}^{~3} & n_f \Big( {5 N_c\over72} + {1\over 18 N_c} \Big) + {1\over36} + {7\over 72 N_c^2}  \\
0 & \big(-{5 N_c \over 6} + {2 \over 9 N_c}\big) n_f  & (Z^{(2)})_{4}^{~3} &  {80 N_c^2 \over 27} - {20 \over 9} + {7 \over 27 N_c^2}  + \big({25 N_c \over27} + {13\over 18 N_c} \big)n_f 
\end{pmatrix} \,.
\nonumber
\end{align}
The computation of entries $(Z_{\hat{\cal O}}^{(2)})_1^{\ 1}$ and $(Z_{\hat{\cal O}}^{(2)})_1^{\ 2}$ were explained above in \eqref{eq:ZO1-mixing}. Other entries can be obtained similarly by considering the renormalization of other form factors. For example, from the form factor ${\cal F}_{\hat{\cal O}_1,\gamma}^{(2)}$ one can compute $(Z_{\hat{\cal O}}^{(2)})_1^{\ 4}$, while using the form factor ${\cal F}_{\hat{\cal O}_4,\gamma}^{(2)}$ one can compute $(Z_{\hat{\cal O}}^{(2)})_4^{\ 2}$ and $(Z_{\hat{\cal O}}^{(2)})_4^{\ 4}$.
In this matrix, $(Z_{\hat{\cal O}}^{(2)})_2^{\ 2}$ matches the result for dimension-4 operator ${\rm tr}(F^2)$ in \cite{Gehrmann:2011aa}. The $N_c^2$ terms of $(Z_{\hat{\cal O}}^{(2)})_1^{\ 1}, \, (Z_{\hat{\cal O}}^{(2)})_1^{\ 2}$ were computed in \cite{Jin:2018fak}. All the other entries are given for the first time to our knowledge.
To determine the matrix elements $(Z^{(l)})_{I}^{~3}$, one needs to compute form factors of $\hat{\cal O}_3$ with four partons, which we leave for future work.

\section{Discussion}
\label{sec:discussion}

The results in the last section provide the complete two-loop QCD corrections to Higgs plus 3-parton amplitudes with dimension-7 operators. They are of phenomenological relevance for the LHC experiments, and provide for the first time the top-mass correction of S-matrix elements for Higgs plus one-jet production at N$^2$LO. Results with full top-mass dependence would require a three-loop computation involving a massive subloop, which is beyond the state of the art.
Our computation relies on a combination of modern on-shell unitarity-cut method and IBP reduction. This strategy can be applied efficiently to the case with higher dimension operators in the Higgs effective action.

The final analytic results take remarkable simple form and exhibit intriguing hidden structures. Below we comment on this in more details. First of all, the maximally transcendental parts take universal forms, generalizing the maximal transcendentality principle (MTP) in two aspects. Firstly, the MTP applies to Higgs and three-parton amplitudes with dimension-seven operators, see also \cite{Brandhuber:2017bkg, Jin:2018fak, Brandhuber:2018xzk, Brandhuber:2018kqb, Jin:2019ile}. Secondly,  the principle applies also to Higgs amplitudes with external quark states, by a change of color factors (see also \cite{Jin:2019ile}):
\begin{align}
& \textrm{Max. Tran. of } {(H \rightarrow q \bar q g)} \big|_{C_F \rightarrow C_A}  
= \textrm{Max. Tran. of } {(H \rightarrow 3g)} \,,
\end{align}
which also match with the maximal transcendental part of the corresponding form factors in ${\cal N}=4$ SYM.
For example, for length-3 operators such as ${\cal O}_1$ and ${\cal O}_4$, the maximal transcendentality part are related to the following universal function:\footnote{This is computed using Catani IR subtraction scheme, and it is different (as purely a scheme difference) from the expression of ${\cal N}=4$ form factors in \cite{Brandhuber:2014ica, Loebbert:2015ova, Brandhuber:2016fni, Loebbert:2016xkw, Brandhuber:2017bkg}, which are based on the BDS subtraction scheme \cite{Bern:2005iz}.}
\begin{align}
R^{(2)}_{\textrm{L3};4}(u,v,w) := &  -{3\over2} {\rm Li}_4(u) + {3\over4} {\rm Li}_4\left(-{u v \over w} \right) - {3\over4} \log(w) \left[ {\rm Li}_3 \left(-{u\over v} \right) + {\rm Li}_3 \left(-{v\over u} \right)  \right] \nonumber\\
& + {\log^2(u) \over 32} \left[ \log^2(u) + \log^2(v) + \log^2(w) - 4\log(v)\log(w) \right] \nonumber\\
& + {\zeta_2 \over 8} \left[ 5\log^2(u) - 2 \log(v)\log(w) \right]- {1\over4} \zeta_4 + \textrm{perms}(u,v,w) \,,
\label{eq:R2L3-def}
\end{align}
and the results in \eqref{eq:R2-O1-alpha-deg4} and \eqref{eq:R2-O4-gamma-deg4} satisfy 
\begin{equation}
{\cal R}^{(2)}_{\hat{\cal O}_1,\alpha;4} =  R^{(2)}_{\textrm{L3};4}(u,v,w) = {\cal R}^{(2), N_c^2}_{\hat{\cal O}_4;\gamma;4} - {\cal R}^{(2), N_c^0}_{\hat{\cal O}_4;\gamma;4} + {\cal R}^{(2), 1/N_c^{-2}}_{\hat{\cal O}_4;\gamma;4} \,.
\label{eq:R2Odeg4-rel}
\end{equation}
Note that the last equality exactly corresponds to taking $C_F \rightarrow C_A$.
Physically, such an identification corresponds to changing the fermions from the fundamental to the adjoint representation. 
This has been known for the kinematic independent quantities such as anomalous dimensions \cite{Kotikov:2002ab, Kotikov:2004er}. For pseudo-scalar Higgs amplitudes involving $q\bar{q}g$ states, the universal maximally transcendental part was also noted in \cite{Banerjee:2017faz}.

For the lower transcendentality parts, the results of QCD and corresponding ${\cal N}=4$ form factors are not identical as expected. Intriguingly, the transcendentality degree-3 and degree-2 parts of QCD results also show some universal structures and have certain connections to the ${\cal N}=4$ results.
In ${\cal N}=4$ form factors, the transcendentality degree-3 part can be expressed in terms of the function $T_3$ \cite{Loebbert:2015ova, Loebbert:2016xkw, Brandhuber:2017bkg}:
\begin{align}
T_3(u,v,w) := & \Big[ -{\rm Li}_3 \left(-{u\over w} \right) + \log(u) {\rm Li}_2\left({v \over 1-u} \right) - {1\over2} \log(1-u) \log\left({w^2\over 1-u}\right) \nonumber\\
& + {1\over2} {\rm Li}_3\left(-{uv \over w}\right) + {1\over2} \log(u)\log(v)\log(w) + {1\over12}\log^3(w) + (u\leftrightarrow v) \Big] \nonumber\\
& +  {\rm Li}_3(1-v) - {\rm Li}_3(u) + {1\over2} \log^2(v) \log\left({1-v\over u}\right) - \zeta_2 \log\left( {u v \over w} \right) \,.
\label{eq:T3-def}
\end{align}
It turns out that the QCD form factors can also be expressed using $T_3$, plus simple $\zeta_3$ or $(\zeta_2\times \log)$ terms, as given in Appendix \ref{app:remainderO1} - \ref{app:remainderO4}. 
We should point out that there are still rational factors associated to the $T_3$ functions which can be different for different form factors, while all the non-trivial transcendental functions are organized together into $T_3$ functions.

For the transcendentality degree-2 parts of QCD form factors, there appear  two main building blocks, $T_2$ and $T'_2$:
\begin{align}
T_2(u,v) := & \text{Li}_2(1-u)+\text{Li}_2(1-v)+\log (u) \log (v)- \zeta_2 \,, 
\label{eq:T2-def} \\
T'_2(u): = &  \text{Li}_2(1-u)+\frac{\log ^2(u)}{2}
\label{eq:T2b-def} \,.
\end{align}
In the case of $r_{\hat{\cal O}_4,\gamma}^{(2),{\rm fin}}$, a few extra ${\rm Li}_2$ functions also exist, but all their coefficients are simple numerical numbers. All  ${\rm Li}_2$ functions with non-trivial rational factors are organized themselves in terms of  $T_2$ and $T'_2$. Other remaining terms are simple $\zeta_2$ and $\log^2$ terms. 
  For the form factor of  ${\cal O}_1\sim{\rm tr}(F^3)$, the $T'_2$ function is not needed, and it was noted that both transcendental degree-2 and degree-1 (log) functions with non-trivial rational kinematic factors are identical between the QCD and ${\cal N}=4$ results \cite{Jin:2018fak}. For example, comparing with \eqref{eq:b5}-\eqref{eq:b6}, the corresponding ${\cal N}=4$ results are\footnote{To compare with the QCD results on an equal footing, the ${\cal N}=4$ results here are also obtained using Catani IR subtraction scheme \cite{Jin:2018fak}, so they are slightly different from those given in \cite{Brandhuber:2017bkg,Brandhuber:2018xzk} using the BDS subtraction scheme \cite{Bern:2005iz}.}
\begin{align}
{\cal R}^{(2),{\cal N}=4}_{{\cal O}_1;2} = & \left(  {u^2\over w^2} -{1\over2} \right)  T_2(u,v)  +3 \log(u) \log(v) -3 \zeta_2  + \textrm{perms}(u,v,w) \,, \\
{\cal R}^{(2),{\cal N}=4}_{{\cal O}_1;1} = & \left( { u^2 \over 2 v w} + {v\over w} -16  \right) \log(u)  + \textrm{perms}(u,v,w) \,. 
\end{align}

As a side comment, we note that in \cite{Badger:2019djh}, a transcendentality-2 building block was found for the two-loop five-gluon double trace scattering amplitudes:
\begin{equation}
I_{123;45}= \text{Li}_2(1-\frac{s_{12}}{s_{123}})+\text{Li}_2(1-\frac{s_{23}}{s_{123}})+\log ^2(\frac{s_{12}}{s_{23}}) + \zeta_2,
\label{1905.03733}
\end{equation}
which is equivalent to the finite part of one-mass box functions. It is similar to our degree-2 building block \eqref{eq:T2-def}. This similarity might be related to the similarity between the kinematics of five-gluon amplitude and three-parton form factors when two gluons with momenta $p_4$ and $p_5$ are merged together.

Knowing the building blocks as described above makes it much easier to simplify the expressions. In particular for the maximal transcendental part, the MTP may allow one to obtain the QCD expression from a much simpler ${\cal N}=4$ result which may be computed to very high loops. One should note that there are also known examples where the maximal transcendentality principle does not apply. For example, MTP does not hold for the four-gluon and five-gluon scattering amplitudes even at one loop.  The one-loop QCD four-gluon amplitudes contain polylogarithm functions, while $\mathcal{N}=4$ SYM amplitudes only contain simple log functions.
Counter examples were also noted in the Regge limit of amplitudes \cite{DelDuca:2017peo} and for the form factor of stress tensor operator \cite{Ahmed:2019nkj}.
By now the sphere of application of MTP is still not clear. It would be interesting to explore the underlying mechanism and consider more examples. Furthermore, it would be important to study further the structures of lower transcendentality parts which are needed to compute full QCD results.
It would be worthy to consider amplitudes in $\mathcal{N}=1,2$ SYM, which may serve as bridges connecting the QCD and $\mathcal{N}=4$ SYM amplitudes.

When scattering amplitudes are classified by transcendental degrees, usually $\frac{1}{s^k}$ type  spurious poles appear. The cancellation of these unphysical poles makes it possible to relate terms with different transcendental degrees, and may be used to constrain the lower transcendentality parts of the amplitude from the higher transcendentality pieces.
The analytical expressions of a subset of two-loop non-planar master integrals for Higgs to 3-parton amplitudes with finite top quark were obtained recently \cite{Bonciani:2019jyb} (which correspond to the NLO order in the Higgs EFT expansion). These integrals contain elliptical sectors. It would be interesting to explore whether there are universal analytical structures in the elliptical sectors.

\acknowledgments

It is a pleasure to thank Lance Dixon, Bo Feng, Hui Luo, Jian-Ping Ma, Ke Ren and Li Lin Yang for discussions. 
We also thank the anonymous referee for useful comments and suggestions.
This work is supported in part by the National Natural Science Foundation of China (Grants No. 11822508, 11947302, 11935013),
by the Chinese Academy of Sciences (CAS) Hundred-Talent Program, 
and by the Key Research Program of Frontier Sciences of CAS. 
We also thank the support of the HPC Cluster of ITP-CAS.

\appendix

\section{One-loop results}
\label{app:oneloop}

In this appendix we provide the one-loop bare form factor results to all order in  $\epsilon$, 
which to our knowledge are given explicitly for the first time in the literature. The higher order terms in $\epsilon$ expansion will be needed in the higher loop computation. $r^{(l)}$ is the normalized form factor defined in \eqref{eq:r-ratioFF-def}. We have:
\begin{align}
r_{b,\hat{\cal O}_1,\alpha}^{(1)} = & u v w N_c {-6 + 10\epsilon - 4 \epsilon^2 - \epsilon^3 \over 2\epsilon (1-\epsilon)(3-2\epsilon) } (I_2[s_{12}]  + I_2[s_{23}] + I_2[s_{13}] ) \,, \\
r_{b,\hat{\cal O}_1,\beta}^{(1)} = & - {v w \over u} N_c {\epsilon^2 \over 2(1-\epsilon)(3-2\epsilon)} I_2[s_{12}] \,, \\
r_{b,\hat{\cal O}_1,\gamma}^{(1)} = & - u N_c \frac{\epsilon }{4 (1-\epsilon) (3-2 \epsilon)} I_2[s_{12}] \,, \\
r_{b,\hat{\cal O}_4,\alpha}^{(1)} = & - u v w n_f {\epsilon\over (1-\epsilon)(3-2\epsilon)} (I_2[s_{12}]  + I_2[s_{23}] + I_2[s_{13}] ) \,, \\
r_{b,\hat{\cal O}_4,\beta}^{(1)} = & - {v w \over u} n_f {\epsilon\over (1-\epsilon)(3-2\epsilon)} I_2[s_{12}] \,, 
\end{align}
\begin{align} %
r_{b,\hat{\cal O}_4,\gamma}^{(1)} = & u \bigg[\frac{ \left(2 \epsilon ^2-\epsilon +2\right)}{4 \epsilon  N_c}-\frac{(1-\epsilon ) n_f}{3-2 \epsilon } \bigg] I_2[s_{12}] \nonumber\\
&+ u \bigg[ \frac{\left(5 \epsilon ^3-9 \epsilon ^2+2 \epsilon \right)}{4 (3-2 \epsilon ) (1-\epsilon ) \epsilon  N_c}+\frac{\left(5 \epsilon ^3-12 \epsilon ^2+11
   \epsilon -6\right) N_c}{4 (3-2 \epsilon ) (1-\epsilon ) \epsilon }  \bigg] I_2[s_{13}] 
 \nonumber\\
& + u \bigg[  \frac{N_c \left(-2 u \epsilon ^2+v \epsilon ^4+v \epsilon ^3-14 v \epsilon ^2+16 v \epsilon -6 v-w \epsilon ^3-w
   \epsilon ^2+\epsilon ^4\right)}{4 v (3-2 \epsilon ) (1-\epsilon ) \epsilon } \nonumber\\
    & \qquad  + \frac{\left(-u \epsilon ^3-u \epsilon ^2+v \epsilon ^4+v \epsilon ^3-10 v \epsilon ^2+6 v \epsilon -2 w \epsilon ^2+\epsilon ^4\right)}{4 v (3-2 \epsilon
   ) (1-\epsilon ) \epsilon  N_c} \bigg] I_2[s_{23}] \,, \\
r_{b,\hat{\cal O}_3,\alpha}^{(1)} = & r_{b,\hat{\cal O}_3,\beta}^{(1)} = r_{b,\hat{\cal O}_3,\gamma}^{(1)} =  0 \,,
\end{align}
where $I_2[s_{ij}]$ are one-loop scalar bubble integrals with $p_i+p_j$ as the external momentum.

\section{Two-loop remainder of $F_{\hat{\cal O}_1}$}
\label{app:remainderO1}

In this and following appendices, we collect the two-loop remainder functions. We follow the definition of $r^{(l)}$ in \eqref{eq:r-ratioFF-def}.

\subsection{$\hat{\cal O}_1 \rightarrow (1^-, 2^-, 3^-)$}

The two-loop finite remainder can be decomposed according to the color factors as:
\begin{equation}
r_{\hat{\cal O}_1,\alpha}^{(2),{\rm fin}} = r_{\hat{\cal O}_1}^{(0)}  \Big( N_c^2\, {\cal R}_{\hat{\cal O}_1,\alpha}^{(2),N_c^2}  + N_c\, n_f\, {\cal R}_{\hat{\cal O}_1,\alpha}^{(2),N_c n_f} + n_f^2\, {\cal R}_{\hat{\cal O}_1,\alpha}^{(2),n_f^2} + {\cal R}^{(2)}_{{\hat{\cal O}_1},\alpha;\log (-q^2)}  \Big) \,,
\end{equation}
in which we separate the terms proportional to $\log(-q^2)$ in ${\cal R}^{(2)}_{{\hat{\cal O}_1},\alpha;\log (q^2)}$.

We decompose the ${\cal R}^{(2),N_c^2}_{\hat{\cal O}_1,\alpha}$ part according to transcendentality degree $d$  as
\begin{align}
{\cal R}^{(2),N_c^2}_{\hat{\cal O}_1,\alpha} = \sum_{d=0}^4  {\cal R}^{(2),N_c^2}_{\hat{\cal O}_1,\alpha;d} \,,
\end{align}
where
\begin{align}
{\cal R}^{(2),N_c^2}_{\hat{\cal O}_1,\alpha;4} = & R^{(2)}_{\textrm{L3};4}(u,v,w)  \,, 
\label{eq:R2-O1-alpha-deg4} \\
{\cal R}^{(2),N_c^2}_{\hat{\cal O}_1,\alpha;3} = & \left( 1+ {u\over w} \right) T_3(u,v,w) + {143\over72}\zeta_3 - {11\over24}\zeta_2\log(u)+ \textrm{perms}(u,v,w) \,, \\
{\cal R}^{(2),N_c^2}_{\hat{\cal O}_1,\alpha;2} = & \left(  {u^2\over w^2} -{1\over2} \right)  T_2(u,v)  - {55 \over 48}\log^2(u) + {73\over 72}\log(u) \log(v) + {23\over6} \zeta_2  + \textrm{perms}(u,v,w) \,, \label{eq:b5}\\
{\cal R}^{(2),N_c^2}_{\hat{\cal O}_1,\alpha;1} = & \left(  { u^2 \over 2 v w}+ {v\over w} + {119\over18} \right) \log(u)  + \textrm{perms}(u,v,w) \,,  \label{eq:b6} \\
{\cal R}^{(2),N_c^2}_{\hat{\cal O}_1,\alpha;0} = & {487\over 72}{1\over u v w} - {14075\over 216}  \,,
\end{align}
and $R^{(2)}_{\textrm{L3};4}(u,v,w)$,  $T_3(u,v,w)$, $T_2(u,v)$ are defined in \eqref{eq:R2L3-def}, \eqref{eq:T3-def} and \eqref{eq:T2-def} respectively.

In \eqref{eq:b5}, there seems to be $\frac{1}{w^2}$-type unphysical poles. Such poles can be cancelled by the zero of $T_2(u,v)$ when $w\rightarrow0 $:
\begin{equation}
T_2(u,1-u-w)=-w\Bigl[\frac{\ln u}{1-u}+\frac{\ln (1-u)}{u}\Bigr]+\mathcal{O}(w^2)\ .
\end{equation}

The $n_f$ parts are simpler and we collect terms of different degrees together:
\begin{align}
{\cal R}_{\hat{\cal O}_1,\alpha}^{(2),N_c n_f} = & \frac{1}{12} \zeta_2 \log (u)-\frac{13 \zeta_3}{36}-\frac{\log ^2(u)}{4}+\frac{1}{18} \log (u) \log (v)-\frac{95 \zeta_2}{72}-\frac{64 \log(u)}{27} +\frac{2863}{648} \nonumber\\  
& + \textrm{perms}(u,v,w) \,, \\
{\cal R}_{\hat{\cal O}_1,\alpha}^{(2),n_f^2} = & \frac{\log ^2(u)}{12}+\frac{1}{18} \log (u) \log (v) +\frac{\zeta_2}{36} -\frac{5 \log (u)}{27}  + \textrm{perms}(u,v,w) \,.
\end{align}

Finally, the terms containing $\log (-q^2)$ are give by:
\begin{align}
{\cal R}^{(2)}_{{\hat{\cal O}_1},\alpha;\log (-q^2)} = & \left(-\frac{19}{24} N_c^2-\frac{7 N_c n_f}{6}+\frac{5 n_f^2}{6}\right) \log ^2(-q^2)  \\
& + \left[ N_c^2 \left(-3 \zeta_3 -\frac{11 \zeta_2}{4}-\frac{19}{36} \log (u v w)-\frac{2}{u v w}+\frac{119}{3}\right)  \right. \nonumber \\
& \quad \left. +N_c n_f
   \left(\frac{\zeta_2}{2}-\frac{7}{9} \log (u v w)-\frac{128}{9}\right) +n_f^2 \left(\frac{5}{9} \log (u v w)-\frac{10}{9}\right)\right] \log (-q^2)   \,.\nonumber
\end{align}


\subsection{$\hat{\cal O}_1 \rightarrow (1^-, 2^-, 3^+)$}

The $(1^-, 2^-, 3^+)$ configuration is very simple:
\begin{align}
r_{\hat{\cal O}_1,\beta}^{(2),{\rm fin}} = & \ N_c^2 \left\{ \Big[ T_2(u,v) + u \log(u)  + \textrm{cyclic perms}(u,v,w) \Big] +  {487\over 72} - {v w \over 36 u}  \right\}  \nonumber\\ & \ + N_c n_f  {v w \over 18 u} - 2 N_c^2 \log(-q^2) \,.
\end{align}

\subsection{$\hat{\cal O}_1 \rightarrow (1^q, 2^{\overline{q}}, 3^-)$}

We decompose the two-loop finite remainder according to the color factors as:
\begin{equation}
r_{\hat{\cal O}_1,\gamma}^{(2),{\rm fin}} = N_c^2\, {\cal R}_{\hat{\cal O}_1,\gamma}^{(2),N_c^2} + N_c^0\, {\cal R}_{\hat{\cal O}_1,\gamma}^{(2),N_c^0} + N_c\, n_f\, {\cal R}_{\hat{\cal O}_1,\gamma}^{(2),N_c n_f} +{\cal R}^{(2)}_{{\hat{\cal O}_1},\gamma;\log (-q^2)}  \,,
\end{equation}
where
\begin{align}
{\cal R}_{\hat{\cal O}_1,\gamma}^{(2),N_c^2} = & \frac{u v}{w} T_2(u,v) +\frac{u w}{v} T_2(u,w)+u \left(\frac{19 \log (u)}{18}+\frac{\log (v)}{18}-\frac{\log (w)}{72}\right) \nonumber\\
& +\frac{u^2}{72 v}+\frac{u}{72 v}-\frac{1501 u}{432}+\frac{487}{72} \,, \\
{\cal R}_{\hat{\cal O}_1,\gamma}^{(2),N_c^0} = & -\frac{v w}{u} T_2(v,w)+\frac{u (-1885 v+6 w+6)}{432 v}+\left(\frac{u}{12}-v\right) \log (v)+\left(\frac{u}{36}-w\right) \log (w)  \,, 
\end{align}
\begin{align} %
{\cal R}_{\hat{\cal O}_1,\gamma}^{(2),N_c n_f} = & \frac{1}{9} u \left(-\log (u)-\frac{\log (v)}{8}-\frac{\log (w)}{8}+\frac{31}{12}\right) \,, \\
{\cal R}^{(2)}_{{\hat{\cal O}_1},\gamma;\log (-q^2)} = & \left[ -\frac{5}{36} u N_c n_f+\left(\frac{79 u}{72}-2\right) N_c^2+\frac{10 u}{9} \right]  \log (-q^2)  \,,
\end{align}
and $T_2(u,v)$ are defined in \eqref{eq:T2-def}.

\section{Two-loop remainder of $F_{\hat{\cal O}_3}$}
\label{app:remainderO3}

For ${\hat{\cal O}_3}$ with three partons, only the $(1^q, 2^{\overline{q}}, 3^-)$-configuration is non-zero. We have
\begin{align}
r_{\hat{\cal O}_3,\gamma}^{(2),{\rm fin}} = & N_c n_f \left(\frac{5 \log
   (v)}{36}-\frac{251}{432}\right) +\frac{n_f}{N_c} \left(\frac{\log (v)}{9}-\frac{43}{108}\right)+\frac{1}{N_c^2} \left( \frac{7 \log
   (v)}{36}-\frac{331}{432} \right) \nonumber\\
& + \left( \frac{\log (v)}{18}-\frac{23}{108} \right) +  \log (-q^2) \left(\frac{5 N_c n_f}{36}+\frac{n_f}{9 N_c}+\frac{7}{36 N_c^2}+\frac{1}{18}\right) \,.
\end{align}

\section{Two-loop remainder of $F_{\hat{\cal O}_4}$}
\label{app:remainderO4}

\subsection{$\hat{\cal O}_4 \rightarrow (1^-, 2^-, 3^-)$}

We decompose the two-loop finite remainder according to the color factors as:
\begin{equation}
r_{\hat{\cal O}_4,\alpha}^{(2),{\rm fin}} = N_c\, n_f\, {\cal R}_{\hat{\cal O}_4,\alpha}^{(2),N_c n_f} + {n_f \over N_c} \, {\cal R}_{\hat{\cal O}_4,\alpha}^{(2),{n_f / N_c}} + n_f^2\, {\cal R}_{\hat{\cal O}_4,\alpha}^{(2),n_f^2} + {\cal R}^{(2)}_{{\hat{\cal O}_4},\alpha;\log (-q^2)} \,,
\end{equation}
where

\begin{align}
{\cal R}_{\hat{\cal O}_4,\alpha}^{(2),N_c n_f} = & \left(\frac{2 u^3 v^3}{w^3}+\frac{5 u^2 v^2}{2 w^2}+\frac{u v}{w}\right) T_2(u,v) \nonumber \\
& +\left(\frac{4 u^3 v^2}{w^2}+\frac{3 u^3 v}{w}+\frac{5 u^2 v^2}{w}+\frac{5 u^3}{12}+\frac{13 u^2 v}{2}+2 u v^2+\frac{109 u v w}{36}\right) \log (u) \nonumber\\
& +\left(\frac{221 u^3}{72}+\frac{81 u^2 v}{4}+\frac{655 u v w}{108}+\frac{u^2 v^2}{w}\right) \,,
\end{align}
\begin{align}
{\cal R}_{\hat{\cal O}_4,\alpha}^{(2),{n_f / N_c}} = & -\frac{u^3 v^3 }{3 w^3} T_2(u,v) +\left(-\frac{2 u^3 v^2}{3 w^2}+\frac{u^3 v}{3
   w}-\frac{u^3}{9}+\frac{u v w}{6}\right) \log (u) \nonumber\\
& +\left(-\frac{139}{216} u^3 -\frac{15 u^2 v}{4}-\frac{55 u v w}{27}-\frac{u^2 v^2}{6 w}\right)  \,, \\
{\cal R}_{\hat{\cal O}_4,\alpha}^{(2),n_f^2} = & - \frac{5}{18}u v w \log (u)+\frac{5 u v w}{27} \,, \\
{\cal R}^{(2)}_{{\hat{\cal O}_4},\alpha;\log (-q^2)} = & \left\{ \left(\frac{4}{9 N_c}-\frac{5 N_c}{3}\right) n_f+\left[\left(\frac{4}{3 N_c}+\frac{25 N_c}{6}\right) n_f-\frac{5
   n_f^2}{3}\right] u v w \right\} \log (-q^2) \,,
\end{align}
and $T_2(u,v)$ are defined in \eqref{eq:T2-def}.

\subsection{$\hat{\cal O}_4 \rightarrow (1^-, 2^-, 3^+)$}

We decompose the two-loop finite remainder according to the color factors as:
\begin{equation}
r_{\hat{\cal O}_4,\beta}^{(2),{\rm fin}} = N_c\, n_f\, {\cal R}_{\hat{\cal O}_4,\beta}^{(2),N_c n_f} +  {n_f \over N_c}\, {\cal R}_{\hat{\cal O}_4,\beta}^{(2),n_f/N_c} + n_f^2\, {\cal R}_{\hat{\cal O}_4,\beta}^{(2),n_f^2} +{\cal R}^{(2)}_{{\hat{\cal O}_4},\beta;\log (-q^2)}  \,,
\end{equation}
where
\begin{align}
{\cal R}_{\hat{\cal O}_4,\beta}^{(2),N_c n_f} = & \frac{2}{3} v T_3(v,u,w)-\left(\frac{2 v^3 w}{3 u^3}+\frac{v^2 w}{u^2}+\frac{v w}{u}\right) T_3(w,v,u) \nonumber\\
& + v w \left(\frac{4 v^2 w^2}{u^5}+\frac{5 v w}{4 u^4}-\frac{35 v w}{12 u^3}+\frac{11}{12 u^3}-\frac{1}{4 u^2}+\frac{7}{6 u}\right) T_2(v,w) \nonumber\\
& +\left(\frac{4 u v^3}{w^3}+\frac{v}{3 u w}-\frac{25 v^3}{6 w^2}+\frac{5 v^2}{w^2}-\frac{13 v^2}{2 w}+\frac{5 v}{3 w}-\frac{7 v}{3}\right) T_2(u,v) \nonumber\\
& +\frac{2 (1-v) v w }{3 u^2} T'_2(v) \nonumber\\
&+\left(\frac{4 u v^2}{w^2}+\frac{5 v w}{6 u}+\frac{15 u}{8}+\frac{1}{6 u}-\frac{13 v^2}{6 w}+\frac{3 v}{w}-\frac{13}{8}\right)\log (u)  \nonumber\\
& + \left(\frac{8 v^3 w^2}{u^4}+\frac{4 v^2 w^2}{u^3}-\frac{3 v^2 w}{2 u^3}+\frac{19 v^3}{6 u^2}-\frac{7 v^2}{4 u^2}+\frac{7 v}{6 u^2}+\frac{37 u^2}{6 w}+\frac{44 w^2}{9 u} \right. \nonumber\\
& \qquad -\frac{383 w}{36 u}-\frac{27 u}{2 w}-\frac{7}{6 u w}+\frac{47 u}{4}+\frac{83}{12 u}+\frac{4 v^3}{w^2}+\frac{5}{3 (v-1)}-\frac{4}{3 (v-1)^2} \nonumber \\
& \left. \qquad +\frac{377 w}{36}+\frac{17}{2 w}-\frac{47}{3}\right) \log (v)  \nonumber\\
& + \left( \frac{2 v^2 w^2}{u^3}-\frac{v w}{24 u^2}-\frac{385 v w}{216 u}-\frac{25 u}{24}+\frac{13}{24 u}+\frac{2 v^2}{w}+\frac{4}{3 (v-1)}+\frac{353}{72} \right) + { (v \leftrightarrow w)}  \,,
\end{align}
\begin{align}
{\cal R}_{\hat{\cal O}_4,\beta}^{(2),{n_f / N_c}} = & \, \left(\frac{2 v^4}{3 w^3}-\frac{2 v^3}{3 w^3}+\frac{2 v^3}{9 w^2}+\frac{4 v}{9}\right) T_2(u,v) \nonumber\\
 & -v w \left(\frac{2 v^2 w^2}{3 u^5}+\frac{2 v w}{3 u^4}-\frac{4 v w}{9 u^3}+\frac{2}{9 u^3}-\frac{1}{9 u^2}+\frac{2}{9 u}\right) T_2(v,w) \nonumber\\
 & + \left(-\frac{u}{18}+\frac{2 w^3}{3 v^2}-\frac{2 w^2}{3 v^2}-\frac{v^2}{9 w}+\frac{v}{3 w}-\frac{1}{18}\right) \log (u) \nonumber\\
 & - \left(\frac{4 v^3 w^2}{3 u^4}+\frac{2 \left(v^2 w^2+v^2 w\right)}{3 u^3}+\frac{2 \left(3 v w^2-2 v w+v\right)}{9 u^2}+\frac{v w}{9 u}+\frac{v}{9 u w}+\frac{2 v}{9 u} \right. \nonumber\\
 & \left. \qquad +\frac{2 v^3}{3 w^2}-\frac{v^2}{9 w}-\frac{v}{9 w}+\frac{v}{9}+\frac{2}{9 (v-1)}-\frac{2}{9 (v-1)^2}+\frac{4}{9}\right) \log (v)  \nonumber\\
 & -\left( \frac{v^2 w^2}{3 u^3}+\frac{2 v w}{9 u^2}+\frac{19 v w}{36 u}+\frac{u}{9}+\frac{1}{6 u}+\frac{w^2}{3 v}+\frac{2}{9 (v-1)}+\frac{127}{216} \right) + { (v \leftrightarrow w)} \,, 
\end{align}
\begin{align} %
{\cal R}_{\hat{\cal O}_4,\beta}^{(2),n_f^2} = & \, \frac{10 v w}{27 u}-\frac{v w (3 \log (u)+\log (v)+\log (w))}{9 u} \,, \\
{\cal R}^{(2)}_{{\hat{\cal O}_4},\beta;\log (-q^2)} = & \left( -\frac{5 N_c n_f (6 u-5 v w)}{18 u}+\frac{4 n_f (u+v w)}{9 u N_c}-\frac{5 v w n_f^2}{9 u} \right) \log (-q^2) \,,
\end{align}
and $T_3(u,v,w)$, $T_2(u,v)$, and $T'_2(u)$ are defined in \eqref{eq:T3-def}, \eqref{eq:T2-def}, and  \eqref{eq:T2b-def}, respectively.

\subsection{$\hat{\cal O}_4 \rightarrow (1^q, 2^{\overline{q}}, 3^-)$}

The two-loop finite remainder can be decomposed according to the color factors as:
\begin{align}
r_{\hat{\cal O}_4,\gamma}^{(2),{\rm fin}} = & N_c^2\, {\cal R}_{\hat{\cal O}_4,\gamma}^{(2),N_c^2} + N_c^0\, {\cal R}_{\hat{\cal O}_4,\gamma}^{(2),N_c^0} + {1\over N_c^2}\, {\cal R}_{\hat{\cal O}_4,\gamma}^{(2),N_c^{-2}}  + {n_f \over N_c}\, {\cal R}_{\hat{\cal O}_4,\gamma}^{(2),n_f/N_c} + N_c\, n_f\, {\cal R}_{\hat{\cal O}_4,\gamma}^{(2),N_c n_f} \nonumber\\ 
& + n_f^2\, {\cal R}_{\hat{\cal O}_4,\gamma}^{(2),n_f^2} + {\cal R}^{(2)}_{{\hat{\cal O}_4},\gamma;\log (-q^2)}  \,.
\end{align}

We further decompose the functions according to transcendentality degree $d$  as
\begin{align}
{\cal R}^{(2)}_{\hat{\cal O}_4,\gamma} = \sum_{d=0}^4  {\cal R}^{(2)}_{\hat{\cal O}_4,\gamma;d} \,.
\end{align}

\noindent
{\bf The degree-4 part:} 
\begin{align}
{\cal R}^{(2), N_c^2}_{\hat{\cal O}_4;\gamma;4} =  & \, 
G(1-v,1-v,1,0,w) -\text{Li}_4(1-v)-\text{Li}_4(v) + \text{Li}_4\Big(\frac{v-1}{v}\Big) +\text{Li}_3(v) \log \Big(\frac{u}{1-v}\Big) \nonumber \\
&
+ \text{Li}_3\Big(\frac{u}{1-v}\Big)\log (v)  + \big[ \text{Li}_3(v)+\text{Li}_3(1-v) \big] \left[ \log (w)-2 \log \Big(\frac{u}{1-v}\Big)\right]
\nonumber   \\ 
& + \text{Li}_2\Big(\frac{u}{1-v}\Big) \text{Li}_2\Big(\frac{v-1}{v}\Big) +\frac{1}{2} \text{Li}_2(v) \log
   ^2\Big(\frac{u}{1-v}\Big)
+\text{Li}_2(1-v) \log (1-v) \log \Big(\frac{u}{1-v}\Big)
\nonumber\\
&+ \frac{1}{24} \log ^2(v)  \left[ \log (v) \log \Big(\frac{v}{(1-v)^4}\Big)-3 \log (w) \log
   \Big(\frac{w}{(1-v)^4}\Big) \right]
\nonumber  \\
& + \zeta_2 \left[ \text{Li}_2(1-v) +\log (v) \log \left(\frac{v}{w}\right) 
- {1\over2} \log ^2\Big(\frac{u}{1-v}\Big) \right] 
 +\zeta_3 \left[ \log \Big(\frac{u}{1-v}\Big)-5 \log (v)\right] 
\nonumber\\
&   +\frac{23 \zeta_4}{8} +  \{v \, \leftrightarrow\, w\} \,,  \nonumber\\
{\cal R}^{(2), 1/N_c^2}_{\hat{\cal O}_4;\gamma;4} = & \, 6 \zeta_3 \log(u) - \frac{11}{2}\zeta_4 \,, \nonumber\\
{\cal R}^{(2), N_c^0}_{\hat{\cal O}_4;\gamma;4} = & \, - \Big( R^{(2)}_{\textrm{L3};4} -{\cal R}^{(2), 1/N_c^2}_{\hat{\cal O}_4;\gamma;4} - {\cal R}^{(2), N_c^2}_{\hat{\cal O}_4;\gamma;4} \Big) \,.
\label{eq:R2-O4-gamma-deg4}
\end{align}
The $n_f$ parts do not contribution to the maximal transcendentality part. 

The above results have been simplified using the symbol method \cite{Goncharov:2010jf}. It turns out that ${\cal R}^{(2), N_c^2}_{\hat{\cal O}_4;\gamma;4}$ part can not be expressed using only classical polylogarithms, which explains the appearance of the multiple polylogarithm function $G(1-v,1-v,1,0,w)$. It is also worth mentioning that the symbol of ${\cal R}^{(2), N_c^2}_{\hat{\cal O}_4;\gamma;4}$ is identical to the universal remainder density of minimal form factors in ${\cal N}=4$ SYM \cite{Brandhuber:2014ica, Loebbert:2015ova, Brandhuber:2016fni, Loebbert:2016xkw}. See \cite{Jin:2019ile} for more discussion on this point.
The above three terms with different color factors, when combined together, reproduce the universal function $R^{(2)}_{\textrm{L3};4}$, as discussed in Section \ref{sec:discussion}. This provides a non-trivial example for the maximal transcendentality principle.

\noindent
{\bf The degree-3 part:} 
\begin{align}
{\cal R}^{(2), N_c^2}_{\hat{\cal O}_4;\gamma;3} =   & \,
\left( 1+ {w\over u} \right)  T_3(v,w,u) + {1\over3} \left( 1+ {v\over u} \right)  T_3(w,v,u) + \zeta_2 \left[ \frac{7 \log (v)}{4} + \frac{\log
   (w)}{12}\right] + \frac{257 \zeta_3 }{18} \,, \\
{\cal R}^{(2), N_c^0}_{\hat{\cal O}_4;\gamma;3} = & \, 
- \left[ \frac{1}{2} \left(\frac{w}{v}+1\right)^2+\frac{w}{v}+1\right] T_3(u,w,v) + T_3(v,u,w) - T_3(v,w,u) \nonumber\\ 
& + \left[ \frac{1}{6} \left(\frac{u}{v}+1\right)^2-\frac{4}{3} \left(\frac{u}{v}+1\right)+\frac{1}{3}\right] T_3(w,u,v) - \frac{1}{3} T_3(w,v,u) \nonumber\\
& - \zeta_2 \left[ \frac{31 \log(u)}{12} - \frac{5 \log (v)}{3} - \frac{7 \log (w)}{6}\right] - \frac{137 \zeta_3 }{36}, \\
{\cal R}^{(2), N_c^{-2}}_{\hat{\cal O}_4;\gamma;3} = & \, 
{w\over u} T_3(v,w,u) + {1\over3} {v\over u}  T_3(w,v,u) + T_3(v,u,w) + {1\over3} T_3(w,u,v) \nonumber\\
& - \frac{\zeta_2}{3} (5 \log (u)+3 \log (v)+\log (w)) - \frac{15 \zeta_3 }{2} \,, 
\end{align}
\begin{align}
{\cal R}^{(2), N_c n_f}_{\hat{\cal O}_4;\gamma;3} =  & \, \frac{2}{3} T_3(u,v,w) + \frac{2}{3} T_3(u,w,v) - \frac{\zeta_2}{6} (8 \log (u)-3 \log (v)-3 \log (w)) - \frac{43 \zeta
   (3)}{9} \,, \\
{\cal R}^{(2), n_f/N_c}_{\hat{\cal O}_4;\gamma;3} =  & \, \frac{\zeta_2}{2} \log (u) - \frac{71 \zeta_3 }{18} \,, \\
{\cal R}^{(2), n_f^2}_{\hat{\cal O}_4;\gamma;3} = & \, 0 \,,
\end{align}
and $T_3(u,v,w)$ are defined in \eqref{eq:T3-def}.

\noindent
{\bf The degree-2 part:} 
\begin{align}
{\cal R}^{(2), N_c^2}_{\hat{\cal O}_4;\gamma;2} =   & \, 
\left(-\frac{v (1-v)^2}{6 u^3}+\frac{(3 v+5) (1-v)}{12 u^2}+\frac{-8 v^2-15 v+6}{18 u v}\right) T_2(v,w) \nonumber\\
& -\frac{13 \text{Li}_2(1-w)}{36}+\frac{8}{9} \log (v)  \log (w)+\frac{17 \log ^2(v)}{18}-\frac{35 \log ^2(w)}{36}+\frac{25 \pi ^2}{18} \,, \\
{\cal R}^{(2), N_c^0}_{\hat{\cal O}_4;\gamma;2} = & \, 
-\left(-\frac{(1-u)^2 u^2}{2 v^4}+\frac{(1-u) (5 u+2) u}{6 v^3}-\frac{7-3 u}{6 v^2}-\frac{8-3 u}{6 v}+\frac{9}{4}\right) T_2(u,w) \nonumber\\
& -\left(-\frac{2 v (1-v)^2}{3 u^3}+\frac{(v+1) (1-v)}{3 u^2}-\frac{39-29 v}{18 u}+\frac{u}{6 v}-\frac{6-23 v}{18 v}\right) T_2(v,w)  \nonumber\\
& +\left(-\frac{5 v^2+v-1}{6 v w}-\frac{v (1-v)^2}{3 w^3}+\frac{(9 v+1) (1-v)}{12 w^2}-\frac{w}{6 v}-\frac{5 v+1}{6 v}\right) T_2(u,v)  \nonumber\\
& +\left(\frac{7}{4}-\frac{u}{6 v}\right) T'_2(w)+\frac{(-u+5 v+1) }{2 v} T'_2(u) \nonumber\\
& +\frac{35 \text{Li}_2(1-v)}{18}+\frac{7 \log ^2(v)}{2}+\frac{7 \log ^2(w)}{12}-\frac{137 \pi ^2}{108} \,, 
\end{align}
\begin{align}
{\cal R}^{(2), N_c^{-2}}_{\hat{\cal O}_4;\gamma;2} = & \, 
-\left(\frac{7 (1-u)^2 u^2}{3 v^4}+\frac{101 u^2-57 u+3}{18 v^2}+\frac{(3-20 u) (1-u) u}{3 v^3}-\frac{2-7 u}{3 v}+1\right) T_2(u,w) \nonumber\\
& +\left(\frac{14 (1-u) u^2}{9 w^3}+\frac{u+6 v-2}{6 v}+\frac{(24-61 u) u}{18 w^2}+\frac{2-17 u}{6 w}\right) T_2(u,v) \nonumber\\
& +\left(-\frac{v (1-v)^2}{u^3}+\frac{(1-v)^2}{3 u^2}-\frac{v^2-6 v+3}{18 u v}-\frac{u}{6 v}+\frac{1-2 v}{3 v}\right) T_2(v,w) \nonumber\\
& -\frac{4 \text{Li}_2(1-v)}{3}-\frac{17 \text{Li}_2(1-w)}{18}-\frac{1}{4} \log^2(w)+\frac{31 \pi ^2}{24} \,, \\
{\cal R}^{(2), N_c n_f}_{\hat{\cal O}_4;\gamma;2} =  & \, 
-\left(\frac{11 (1-u)^2 u^2}{6 v^4}+\frac{131 u^2-90 u+6}{18 v^2}+\frac{(5-18 u) (1-u) u}{3 v^3}-\frac{5 (1-3 u)}{3 v}+\frac{16}{9}\right)
   T_2(u,w) \nonumber\\
& +\left(\frac{11 v (1-v)^2}{9 w^3}-\frac{(v+3) (1-v)}{9 w^2}+\frac{2 v}{9 w}-\frac{4}{9}\right) T_2(u,v)-\frac{1}{12} T_2(v,w)-\frac{10}{9}  \text{Li}_2(1-u) \nonumber\\
& +\frac{19}{36} \text{Li}_2(1-v) -\frac{\text{Li}_2(1-w)}{36}-\frac{35}{18} \log ^2(u)-\frac{23}{36} \log ^2(v)-\frac{2 }{9}\log^2(w)-\frac{11 \pi ^2}{27} \,, 
\end{align}
\begin{align}
{\cal R}^{(2), n_f/N_c}_{\hat{\cal O}_4;\gamma;2} =  & \, 
\left(\frac{2 (2 w+1)}{9 u}-\frac{4}{9}\right) T_2(v,w)+\frac{8 \text{Li}_2(1-u)}{9}+\frac{8 \text{Li}_2(1-v)}{9}+\frac{4 \text{Li}_2(1-w)}{9} \nonumber\\
&-\frac{1}{2} \log ^2(v)-\frac{\log ^2(w)}{6}-\frac{25 \pi ^2}{108} \,, \\
{\cal R}^{(2), n_f^2}_{\hat{\cal O}_4;\gamma;2} = & \, \frac{1}{9} \log (u) \log (v)+\frac{1}{9} \log (u) \log (w)+\frac{4 \log ^2(u)}{9}+\frac{1}{36} \log (v) \log (w)+\frac{5 \log ^2(v)}{72}\nonumber\\
&+\frac{5 \log^2(w)}{72}+\frac{\pi ^2}{108} \,,
\end{align}
where $T_2(u,v)$ and $T'_2(u)$ are defined in \eqref{eq:T2-def} and \eqref{eq:T2b-def}.

\noindent
{\bf The degree-1 part:} 
\begin{align}
{\cal R}^{(2), N_c^2}_{\hat{\cal O}_4;\gamma;1} =   & \, 
\left(\frac{v^2}{6 u^2}-\frac{v}{6 u^2}+\frac{v}{6 u}+\frac{13 u}{18 v}+\frac{5}{12 u}+\frac{19}{18 v}+\frac{101}{12}\right) \log (v) \nonumber\\ 
& + \left(-\frac{w^2}{6 u^2}-\frac{w^2}{6 u v}+\frac{2 w}{3 u v}-\frac{u}{6 v}-\frac{w}{3 u}-\frac{11 w}{36 v}-\frac{2}{9 v (1-w)}+\frac{5}{18
   v}+\frac{391}{24}\right) \log (w) \,, \\
{\cal R}^{(2), N_c^0}_{\hat{\cal O}_4;\gamma;1} = & \, 
\left(\frac{u^2 w}{2 v^3}-\frac{5 u^2}{12 v^2}+\frac{u^2}{4 v w}-\frac{u^2}{3 w^2}-\frac{u w}{3 v^2}+\frac{3 u}{4 v}+\frac{u}{6 w}+\frac{w}{4 v}-\frac{2651}{216}\right)\log (u)   \nonumber\\ 
& +\left(\frac{u^2}{3 w^2}-\frac{2 w^2}{3 u^2}+\frac{2 w}{3 u^2}-\frac{u}{3 w^2}+\frac{u}{12 w}-\frac{4 w}{3 u}+\frac{1}{3 u}+\frac{29 w}{36 v}+\frac{8}{9 v}+\frac{1}{3 w}-\frac{247}{27}\right) \log (v)   \nonumber\\ 
& +\left(\frac{2 w^2}{3 u^2}+\frac{11 u w^2}{12 v^3}+\frac{u w}{12 v^2}-\frac{2 w}{3 u v}+\frac{4 w}{3 u}+\frac{5 w^3}{12 v^3}-\frac{5 w^2}{12 v^3}+\frac{w^2}{12 v^2}-\frac{w}{4 v^2} \right. \nonumber \\
& \qquad \left. +\frac{13}{9 v (1-w)}-\frac{4}{3 v}+\frac{733}{216}\right) \log (w)  \,, 
\end{align}
\begin{align}
{\cal R}^{(2), N_c^{-2}}_{\hat{\cal O}_4;\gamma;1} = & \, 
\left(-\frac{7 u^2 w}{3 v^3}+\frac{13 u^2}{6 v^2}-\frac{23 u^2}{18 v w}+\frac{14 u^2}{9 w^2}-\frac{u w}{v^2}-\frac{u}{12 v}+\frac{4 u}{3 w}-\frac{w}{2 v}-\frac{5}{8}\right) \log (u)  \nonumber\\ 
& +  \left(-\frac{14 u^2}{9 w^2}+\frac{w^2}{u^2}-\frac{w}{u^2}+\frac{14 u}{9 w^2}-\frac{47 u}{18 w}+\frac{17 w}{6 u}-\frac{3}{2 u}-\frac{w}{12 v}-\frac{1}{12 v}+\frac{5}{9 w}-\frac{40}{9}\right)  \log (v) \nonumber\\ 
& + \left(-\frac{w^2}{u^2}-\frac{3 u w^2}{2 v^3}+\frac{u w}{v^2}+\frac{5 w^2}{6 u v}+\frac{5 u}{6 v}-\frac{2 w}{u}+\frac{5 w^3}{6 v^3}-\frac{5 w^2}{6 v^3}+\frac{w^2}{v^2}+\frac{5 w}{3 v}  \right. \nonumber \\
& \qquad \left. -\frac{11}{9 v (1-w)}+\frac{10}{9 v}-\frac{73}{36}\right) \log (w) \,, \\
{\cal R}^{(2), N_c n_f}_{\hat{\cal O}_4;\gamma;1} =  & \, 
\left(\frac{11 u^3}{6 v^3}-\frac{11 u^2}{6 v^3}+\frac{61 u^2}{12 v^2}-\frac{5 u}{3 v^2}+\frac{43 u}{9 v}-\frac{3}{1-u}-\frac{4}{3 (1-u)^2}+\frac{11
   v^2}{9 w^2}-\frac{22 v}{9 w^2}-\frac{v}{2 w} \right. \nonumber \\
& \qquad \left. -\frac{17}{18 v w}+\frac{11}{18 v}+\frac{11}{9 w^2}+\frac{13}{9 w}+\frac{154}{27}\right) \log (u)   \nonumber\\ 
& +\left(-\frac{11
   v^2}{9 w^2}+\frac{11 v}{9 w^2}+\frac{v}{2 w}+\frac{5 w}{36 v}-\frac{5}{18 v}-\frac{1}{3 w}+\frac{7}{108}\right) \log (v) \nonumber\\ 
& +\left(\frac{11 w^3}{6
   v^3}-\frac{11 w^2}{6 v^3}+\frac{5 w^2}{12 v^2}+\frac{2 w}{3 v^2}-\frac{9 w^2 - 11 w+6}{36 v (1-w)} -\frac{223}{54}\right)
   \log (w) \,, 
   \end{align}
\begin{align}
{\cal R}^{(2), n_f/N_c}_{\hat{\cal O}_4;\gamma;1} =  & \, 
\left(\frac{4}{9 (1-u)}+\frac{2}{9 (1-u)^2}+\frac{269}{54}\right) \log (u)-\left(\frac{1-5 w^2}{36 v (1-w)}-\frac{119}{54}\right) \log (w) \nonumber\\ 
& +\left(-\frac{5 w}{36 v}-\frac{5}{36 v}+\frac{275}{54}\right) \log (v)  \,, \\
{\cal R}^{(2), n_f^2}_{\hat{\cal O}_4;\gamma;1} = & \, 
-\frac{40 \log (u)}{27}-\frac{10 \log (v)}{27}-\frac{10 \log (w)}{27} \,.
\end{align}

\noindent
{\bf The degree-0 part:} 
\begin{align}
{\cal R}^{(2), N_c^2}_{\hat{\cal O}_4;\gamma;0} =   & \, 
-\frac{w}{12 u}+\frac{143 w}{72 v}-\frac{3}{v}-\frac{44245}{864} \,, \\
{\cal R}^{(2), N_c^0}_{\hat{\cal O}_4;\gamma;0} = & \, 
-\frac{v}{3 u}+\frac{1}{3 u}-\frac{w^2}{4 v^2}+\frac{w}{4 v^2}-\frac{25 w}{36 v}+\frac{v}{6 w}-\frac{301}{72 v}-\frac{1}{6 w}+\frac{5851}{324}   \,, \\
{\cal R}^{(2), N_c^{-2}}_{\hat{\cal O}_4;\gamma;0} = & \, 
 \frac{13 u w}{24 v^2}-\frac{w}{2 u}+\frac{7 u}{9 w}+\frac{41 w^2}{24 v^2}-\frac{41 w}{24 v^2}-\frac{9}{8 v}+\frac{32303}{2592} \,, \\
{\cal R}^{(2), N_c n_f}_{\hat{\cal O}_4;\gamma;0} =  & \, 
-\frac{4}{3 (1-u)}+\frac{221}{36 u}+\frac{11 w^2}{12 v^2}-\frac{11 w}{12 v^2}-\frac{4 w}{3 v}-\frac{11 v}{18 w}+\frac{85}{36 v}+\frac{11}{18 w}+\frac{4435}{648} \,, \\
{\cal R}^{(2), n_f/N_c}_{\hat{\cal O}_4;\gamma;0} =  & \, 
-\frac{2 u}{3 v}+\frac{2}{9 (1-u)}-\frac{139}{108 u}+\frac{4}{3 v}-\frac{11273}{648}  \,, \\
{\cal R}^{(2), n_f^2}_{\hat{\cal O}_4;\gamma;0} = & \, 
\frac{100}{81} \,.
\end{align}

Finally, the terms containing $\log (-q^2)$ are give by:
\begin{align}
{\cal R}^{(2)}_{{\hat{\cal O}_4},\gamma;\log (-q^2)} = & \left[ N_c^2 \left(-\frac{25 w}{36 v}+\frac{25}{18 v}+\frac{25 \log (v)}{9}-\frac{25 \log (w)}{36}-8 \zeta (3)+\frac{11 \zeta_2}{6}+\frac{1739}{72}\right)  \right. \nonumber\\
& \left. +\frac{1}{N_c^2} \left( \frac{2 w}{9 v}+\frac{2}{9 v}+\frac{4 \log (v)}{3}+\frac{4 \log (w)}{9}+6  \zeta (3)- 3 \zeta_2 -\frac{167}{24} \right)  \right. \nonumber\\
 & 
\left. + N_c n_f \left(-\frac{5}{3 u}-\frac{25 \log (u)}{9}+\frac{5 w}{18 v}-\frac{5}{9 v}-\frac{65 \log (v)}{36}-\frac{5 \log (w)}{12}-\frac{\zeta_2}{3}+\frac{1}{9}\right) \right. \nonumber\\
& \left. +\frac{n_f}{N_c} \left(\frac{4}{9 u}-\frac{8 \log (u)}{9}-\frac{5 w}{18 v}-\frac{5}{18 v}-\frac{17 \log (v)}{9}-\frac{7 \log (w)}{9}+\frac{\zeta_2}{2}+\frac{221}{18}\right) \right. \nonumber\\
& \left. +n_f^2 \left(\frac{10 \log (u)}{9}+\frac{5 \log (v)}{18}+\frac{5 \log (w)}{18}-\frac{20}{9}\right)  \right. \nonumber\\
& \left. + \left( \zeta (3)+\frac{\zeta_2}{4}+\frac{17 w}{36 v}+\frac{41}{36 v}+\frac{91 \log (v)}{18}+\frac{7 \log (w)}{6}-\frac{341}{18} \right) \right] \log (-q^2) \nonumber\\
&  + \left(-\frac{5 N_c n_f}{2}-\frac{16 n_f}{9 N_c}+\frac{25 N_c^2}{24}+\frac{8}{9 N_c^2}+\frac{5 n_f^2}{6}+\frac{28}{9}\right) \log ^2(-q^2) \,.
\end{align}

\providecommand{\href}[2]{#2}\begingroup\raggedright\endgroup

\end{document}